\documentclass[reprint,superscriptaddress,amsmath,assume,aps]{revtex4-2}
\usepackage{graphicx}
\usepackage{dcolumn}
\usepackage{bm}
\usepackage{dcolumn}
\usepackage{color}
\usepackage{siunitx}

\let\maketitlesup\maketitle

\begin{document}

\title{Ferroaxial order of the monolayer ice in martyite}
\author{T. Nomura}
\email{nomura.toshihiro@shizuoka.ac.jp}
\affiliation{Department of Physics, Faculty of Science, Shizuoka University, Shizuoka 422-8529, Japan}
\author{S. Kitou}
\email{kitou@edu.k.u-tokyo.ac.jp}
\affiliation{Department of Advanced Materials Science, The University of Tokyo, Kashiwa 277-8561, Japan}
\author{J. Komatsu}
\affiliation{Department of Chemistry, Faculty of Science, Okayama University, Okayama 700-8530, Japan}
\author{K. Koga}
\email{koga@okayama-u.ac.jp}
\affiliation{Department of Chemistry, Faculty of Science, Okayama University, Okayama 700-8530, Japan}
\affiliation{Research Institute for Interdisciplinary Science, Okayama University, Okayama 700-8530, Japan}
\author{T. Hasegawa}
\affiliation{Graduate School of Advanced Science and Engineering, Hiroshima University, Higashi-Hiroshima 739-8521 739-8521, Japan}
\author{N. Ogita}
\affiliation{Graduate School of Advanced Science and Engineering, Hiroshima University, Higashi-Hiroshima 739-8521 739-8521, Japan}
\author{Y. Nakamura}
\affiliation{Japan Synchrotron Radiation Research Institute, Sayo 679-5198, Japan}
\author{H. Ishikawa}
\affiliation{Institute for Solid State Physics, The University of Tokyo, Kashiwa 277-8581, Japan}
\affiliation{Department of Applied Chemistry, Faculty of Science Division 1, Tokyo University of Science, Tokyo 162-0826, Japan}
\author{T. Yajima}
\affiliation{Institute for Solid State Physics, The University of Tokyo, Kashiwa 277-8581, Japan}
\affiliation{Graduate School of Engineering, Nagoya University, Nagoya 464‑8603, Japan}
\author{A. Matsuo}
\affiliation{Institute for Solid State Physics, The University of Tokyo, Kashiwa 277-8581, Japan}
\author{M. Kofu}
\affiliation{Institute for Solid State Physics, The University of Tokyo, Kashiwa 277-8581, Japan}
\author{O. Yamamuro}
\affiliation{Institute for Solid State Physics, The University of Tokyo, Kashiwa 277-8581, Japan}
\author{Z. Hiroi}
\affiliation{Institute for Solid State Physics, The University of Tokyo, Kashiwa 277-8581, Japan}
\author{Y. Tomita}
\affiliation{College of Engineering, Shibaura Institute of Technology, Saitama 337-8570, Japan}
\author{T. Arima}
\affiliation{Department of Advanced Materials Science, The University of Tokyo, Kashiwa 277-8561, Japan}
\affiliation{RIKEN Center for Emergent Matter Science (CEMS), Wako 351-0198, Japan}
\author{T. Matsuo}
\affiliation{Department of Chemistry and Research Center for Thermal and Entropy Science, Osaka University, Osaka 560-0043, Japan}

\date{\today}

\begin{abstract}
Ice Ih, the most stable phase of water at ambient pressure, is a stacking of the honeycomb network of water molecules H$_2$O.
What if one layer of ice is exfoliated and confined to a two-dimensional (2D) sheet?
Martyite Zn$_3$(V$_2$O$_7$)(OH)$_2$$\cdot$2H$_2$O, a mineral with the honeycomb lattice of H$_2$O in the porous framework, is an ideal system for studying such monolayer ice.
Due to the geometrical frustration and 2D nature, H$_2$O molecules are dynamically disordered at room temperature.
In this study, we reveal disorder-order transitions of H$_2$O in martyite using single-crystal x-ray diffraction (XRD).
The XRD results visualize the formation of hydrogen-bonded toroidal H$_2$O hexamers, leading to the ferroaxial order below 200~K.
Combined with the molecular dynamics simulations, we discuss the formation process of the H$_2$O hexamers and how they compromise the molecular arrangement towards lower temperatures.
Our results unveil the ground state of monolayer ice, a fundamental knowledge to understand the polymorphism of H$_2$O.

\end{abstract}
\maketitle


The polymorphism of water is a central interest of science.
Ice shows diverse forms in different environments, from fluttering snowflakes to dense ice inside the core of planets \cite{RevModPhys.84.885,REDMER2011798, Tschauner2018}.
The natural ice on the Earth (ice Ih) is a hydrogen-bonded crystal with a stack of buckled honeycomb layers of H$_2$O molecules.
The local configuration satisfies the two-covalent and two-hydrogen bonds in the tetrahedral geometry, the so-called ice rules (Bernal–Fowler rules) \cite{Bernal}.
The ice rules allow for multiple proton configurations, causing disorder and residual entropy \cite{Pauling} in ice Ih and other ice forms, which manifests as an orientational disorder. 
Driven by a balance of intermolecular interactions, pressure-volume work, and entropy (including residual entropy), at least 20 ice phases have been reported across various temperature and pressure ranges \cite{Salzmann2019, Millot2019, Salzmann2021, Gasser2021,Falenty2014, Rosso2016, JPCC2016}.

Confining water molecules offers a key strategy for accessing novel water states \cite{Zhao2014, Cui2023}, including the emergence of new kinds of phase transitions like liquid-solid critical points \cite{doi:10.1073/pnas.1422829112}. 
This approach has been used to study interfacial water on metal surfaces and graphene sheets 
\cite{Shiotari2017, Tian2022, Kimmel2009, Algara-Siller2015}, 
where local structures resemble ice Ih (satisfying the ice rule), though the substrate's periodicity globally influences the symmetry. 
Confinement within minerals 
\cite{Winkler1996,Bergman2000,Kolesnikov2016,Gorshunov2016,Belyanchikov2020,MATSUO19731829}, 
molecular-organic frameworks 
\cite{Rieth2019,Fischer2020,Hanikel2021}, 
and carbon nanotubes \cite{Koga2001,Kolesnikov2004,doi:10.1073/pnas.0707917105} also allows for control over intermolecular distance and coordination geometry, expanding the diversity of water structures and phase behavior.

\begin{figure}[bt]
\centering
\includegraphics[width=0.99\linewidth]{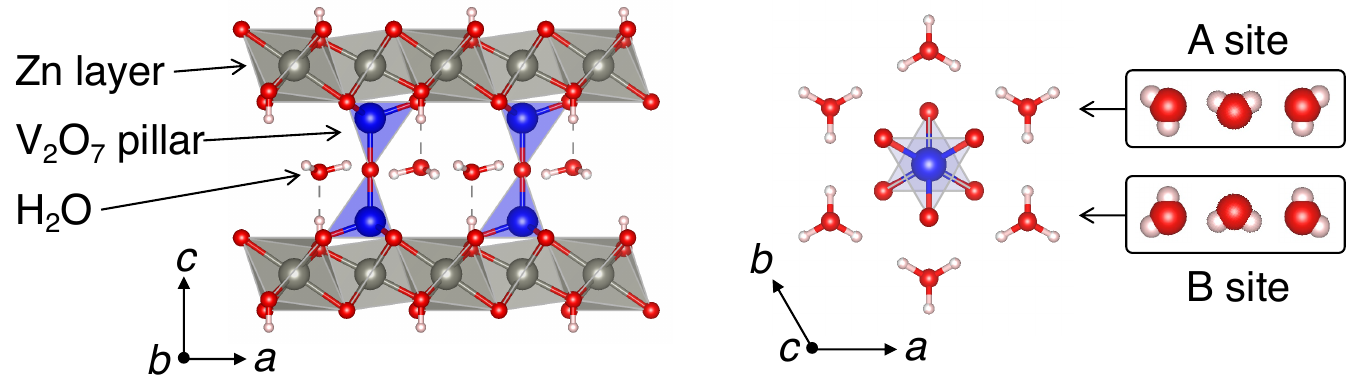}
\caption{
Crystal structure of martyite at room temperature.
Three favored orientations of H$_2$O are shown for the A and B sublattices of the honeycomb lattice.
Dashed lines show the hydrogen bond.
} 
\label{fig1}
\end{figure}

In this study, we focus on a hydrated mineral, martyite Zn$_3$(V$_2$O$_7$)(OH)$_2$$\cdot$2H$_2$O, with the trigonal crystal system (space group $P\overline{3}m1$) \cite{Kampf, Zavalij}.
Martyite has the quasi-2D porous framework, made of the block layers of ZnO$_4$(OH)$_2$ octahedra separated by the V$_2$O$_7$ pillars (Fig.~\ref{fig1}).
The quasi-2D water of crystallization forms the honeycomb network of H$_2$O, which is isostructural to the monolayer ice.
Each H$_2$O is coordinated by the hydroxy group of the framework and forms the hydrogen bonds alternately in the $\pm c$ directions.
This hydrogen bond moderately fixes the oxygen atom of H$_2$O, while, the proton configuration (molecular orientation) can change in the $ab$ plane.
As a result, the H$_2$O molecules in martyite are dynamically disordered with the three favored orientations along the nearest-neighbor directions of the honeycomb network, similar to the plastic ice VII \cite{Rescigno2025}.

Our central question is, how these H$_2$O molecules are arranged at low temperatures.
As a first approximation, considering the three orientations at the A and B sites of the honeycomb lattice, four types of AB-site configurations are expected, and only one of them can form hydrogen bond (see supplemental materials, SM \cite{suppl}).
We comment that the out-of-plane interaction energy is of the order of 1~\% of the in-plane interaction (see SM \cite{suppl}).
Thus, water in martyite can be regarded as monolayer ice (quasi-2D system), where the in-plane hydrogen bonding plays a dominant role.
On the honeycomb lattice of H$_2$O, the average number of hydrogen atoms per bond is 4/3.
Namely, all the nearest-neighbor molecules cannot be hydrogen-bonded.
Therefore, the orientational degree of freedom of the monolayer ice is geometrically frustrated, leading to a highly degenerate ground state.
In this study, we reveal the ground state of monolayer ice and discuss how the system lifts the frustration at low temperatures.

We prepared single- and polycrystals of martyite by hydrothermal synthesis. 
We measured the dielectric properties of polycrystalline martyite pellets by using an LCR meter.
We performed the x-ray diffraction (XRD) experiments using a single crystal in the temperature range from 30 to 300~K on BL02B1 at a synchrotron facility SPring-8 \cite{10.1063/1.3463359}. 
We carried out molecular dynamics (MD) simulations of two-dimensional water in martyite using GROMACS 2018 software \cite{abraham2015softwarex} under periodic boundary conditions in the two dimensions. 
Experimental details are described in SM \cite{suppl}.

\begin{figure}[tb]
\centering
\includegraphics[width=0.8\linewidth]{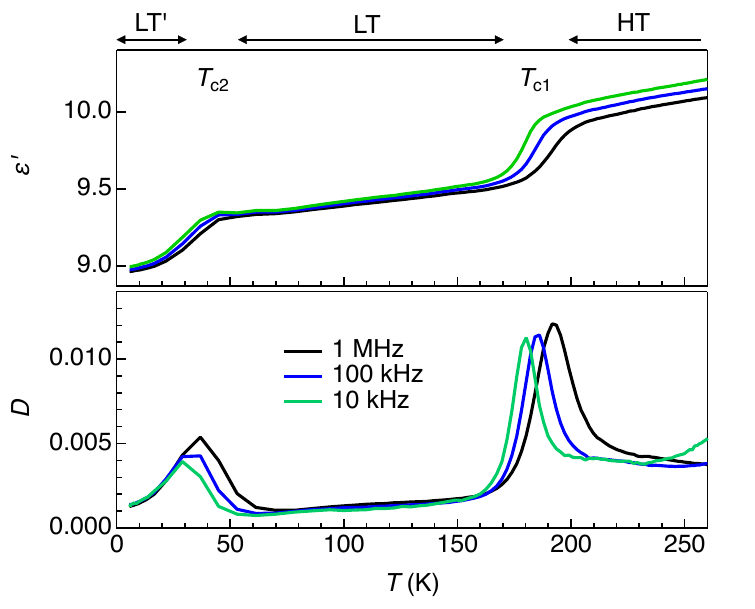}
\caption{
Temperature dependence of the dielectric constant $\varepsilon'$ and dissipation factor $D$ of the polycrystalline martyite. 
} 
\label{dielectric}
\end{figure}

Figure~\ref{dielectric} shows the dielectric constant $\varepsilon'$ and dissipation factor $D$ of the polycrystalline martyite as a function of temperature $T$.
Two anomalies are observed at $T_\mathrm{c1} \approx$ 170--200~K and $T_\mathrm{c2} \approx$ 30--50~K.
The decreases of $\varepsilon'$ indicate the ordering of the H$_2$O molecules rotating at room temperature.
Based on the XRD, Raman spectra, and MD simulation, we conclude that these anomalies are related to the two structural phase transitions driven by the disorder-order transition of H$_2$O (see following sections and Raman spectra data in SM \cite{suppl}).
Accordingly, we term the phases as HT ($T_\mathrm{c1}<T$), LT ($T_\mathrm{c2}<T<T_\mathrm{c1}$), and LT' ($T<T_\mathrm{c2}$).
The frequency dependences of $T_\mathrm{c1}$ and $T_\mathrm{c2}$ indicate the dielectric relaxation where the orientational degree of freedom gradually freezes.
The observed dynamics reflect the energy barrier for rotational motions of H$_2$O to relax to nearly-degenerate but lower energy states.

\begin{figure}[bt]
\centering
\includegraphics[width=0.99\linewidth]{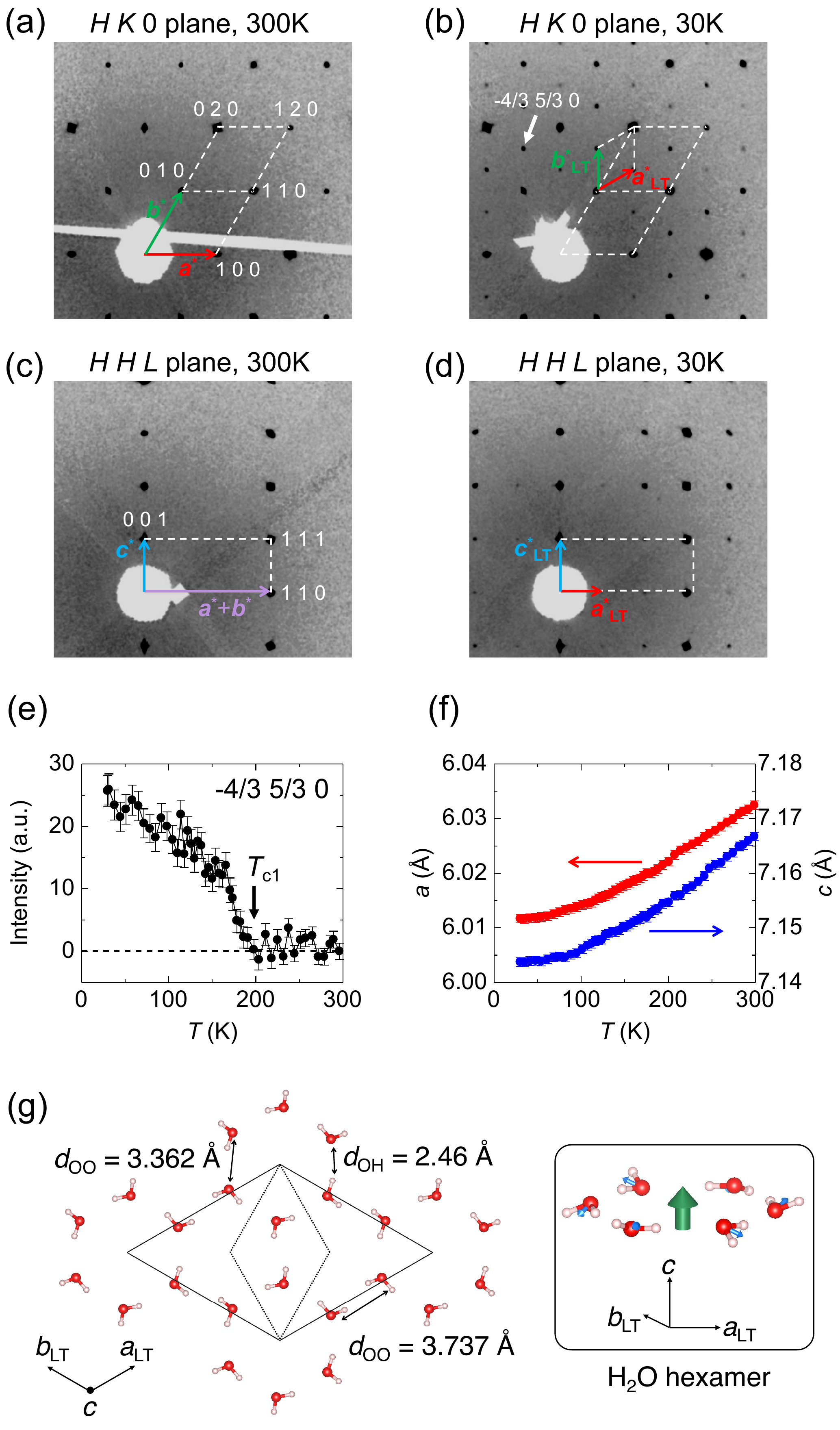}
\caption{
(a-d) XRD patterns of the single crystal of martyite at (a, c) 300~K and (b, d) 30~K.
(e) Temperature dependence of the superlattice Bragg peak intensity at the index of $-4/3$, $5/3$, 0.
(f) Temperature dependence of the lattice constants $a$ and $c$ of the original unit cell.
(g) Ordered structure of H$_2$O molecules in martyite at 30~K. 
The dotted and solid lines show the HT and LT unit cells, respectively.
A perspective view of one H$_2$O hexamer is shown in the panel.
The blue and green arrows show electric dipole moments of H$_2$O and electric toroidal dipole moments of H$_2$O hexamers, respectively.
} 
\label{XRD}
\end{figure}

Figures~\ref{XRD}(a-d) show the single-crystal XRD patterns of martyite at 300~K and 30~K. 
We analyzed the XRD data at 300 K and confirmed that the crystal structure is consistent with the previous reports \cite{Kampf, Zavalij} (see SM \cite{suppl}). 
Between the layers of martyite, the O atoms of H$_2$O molecules form a slightly buckled honeycomb network with equivalent O-O bonds.
Each O atom is bonded to three H atoms with a 2/3 occupancy, allowing the H$_2$O molecule to retain the orientation degree of freedom of the three-state vector (Fig.~\ref{fig1}).
Superlattice peaks appear at the wave vector $q=\lbrace1/3, 1/3, 0\rbrace$ below $T_\mathrm{c1}\approx 200$~K, indicating a structural phase transition.
Figure~\ref{XRD}(e) shows the temperature dependence of the superlattice peak intensity -4/3 5/3 0 as an order parameter.
No peak splitting or extinction rules are observed in any of the fundamental and superlattice peaks. 
Therefore, the unit cell becomes $\sqrt{3}a \times \sqrt{3}a \times c$, and the trigonal symmetry is maintained. 
We note that no superlattice peaks or diffuse scattering are observed along the $c^*$-axis in the LT phase. 
This indicates that the ordering of H$_2$O molecules is in-phase along the $c$ axis and is long-range ordered in martyite, in contrast to the proton-disordered ice Ih \cite{Keen2015}.

We conclude that the space group of the LT phase at 30~K is $P\overline{3}$ (see SM \cite{suppl}).
This symmetry lowering generates merohedral twins due to the disappearance of mirror planes perpendicular to the tertiary axes of the trigonal lattice.
In the current crystal, these domains exist in an approximately 1:1 ratio (see SM \cite{suppl}). 
Figure~\ref{XRD}(g) shows the ordered arrangement of H$_2$O molecules at 30~K. 
The O atoms of the H$_2$O molecules, which form a honeycomb lattice in the HT phase, are distorted into a breathing honeycomb lattice, forming H$_2$O hexamers in the LT phase. 
In this hexamer, H$_2$O molecules form a toroidal ring.
One H atom in the H$_2$O molecule points toward a neighboring H$_2$O molecule, while the other H atom is offset from the honeycomb plane. 
The off-plane H atoms are arranged alternately up and down as coordinated from the ZnO$_4$(OH)$_2$ block layer.
This alternation in the monolayer ice resembles the hydrogen bonds in ice Ih formed in the $\pm c$ directions.

Figure~\ref{XRD}(f) shows the temperature dependence of the lattice constants. 
There is no anomaly in the $a$ and $c$ parameters, suggesting that the framework of martyite is rigid against the breathing deformation of the honeycomb network of H$_2$O.
The observed superlattice peak intensities are bilinear to the atomic displacements of O of H$_2$O (see SM \cite{suppl}).

\begin{figure*}[tb]
\centering
\includegraphics[width=0.9\linewidth]{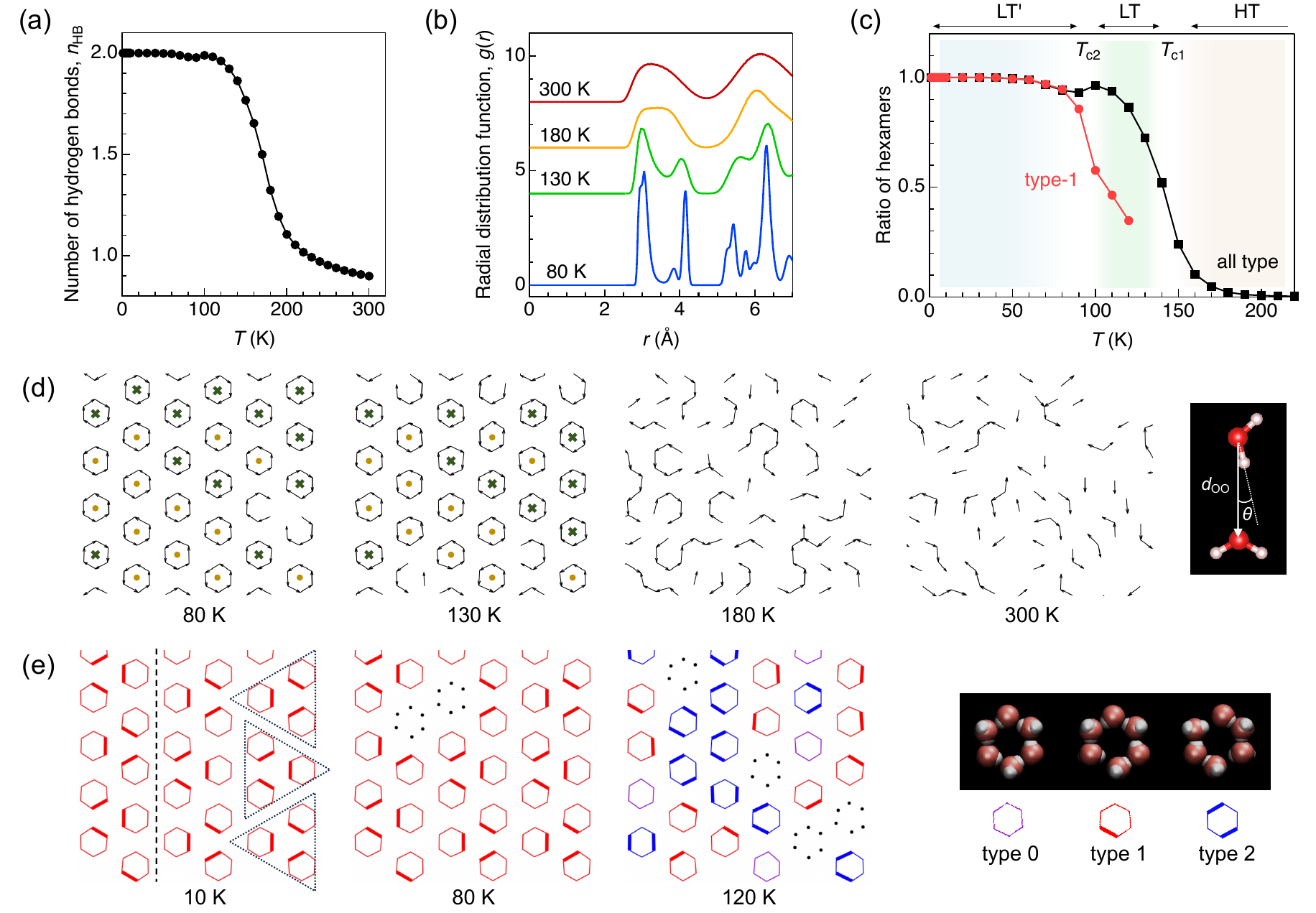}
\caption{
(a) The number of hydrogen bonds per water molecule as a function of temperature. 
(b) The in-plane radial distribution function between O atoms of H$_2$O. The curves are shifted for clarity. 
(c) The ratio of the hexamers formed on the honeycomb lattice. 
The black and red symbols represent all types of hexamers and type-1 hexamers, respectively.
The ratio of the type-1 hexamers is calculated for the designed ferrotoroidal state.
(d) The snapshots of the hydrogen bonding network. 
The arrows represent the hydrogen-bonded pair defined by the intermolecular O--O distance ($d_\mathrm{OO}$) and the hydrogen bonding angle ($\theta$) in the right figure.
Clockwise (counterclockwise) hexamers are marked by green crosses (yellow circles).
(e) Spatial distribution of the hexamers, type 0, 1, and 2, simulated for the designed ferrotoroidal state.
As shown in the right figures, purple (0), red (1), and blue (2) hexagons represent the H$_2$O hexamers with zero, one, and two longer hydrogen bonds.
The black dots indicate that at least one hydrogen bond in the hexamer is broken.
The dotted triangles indicate the H$_2$O octadecamers.
The dashed line indicates the domain boundary of the lattice of the octadecamers.
}
\label{MDfig01}
\end{figure*}

MD simulations on the monolayer ice in martyite demonstrate how the randomly oriented water molecules start to form hydrogen bonds with their neighbors. 
The simulation movie is given in SM \cite{suppl}. 
In the simulation, we use the water model TIP4P/ICE \cite{10.1063/1.1931662} and design one layer of martyite, namely, the monolayer ice sandwiched by the ZnO$_4$(OH)$_2$ layers supported by the V$_2$O$_7$ pillars, where the structural parameters are taken from the XRD results.
Water molecules interact with each other and with atoms of the framework via Lennard-Jones and Coulomb potentials.
Although no artificial constraints are imposed on the water molecules, their translational motions are naturally restricted to those around a honeycomb lattice site even at 300~K.
No quantum effect on the dynamics of atoms is assumed in the MD simulations.

Figure~\ref{MDfig01}(a) shows the number of hydrogen bonds per water molecule $n_{\rm HB}$ as a function of temperature cooled down from 300~K.
The results obtained with the TIP4P/2005 model are presented in SM \cite{suppl}, which are essentially the same as those with the TIP4P/ICE model.
At 300~K, $n_{\rm HB} < 1$, meaning that each water molecule in martyite assumes an almost random orientation. 
As the temperature decreases, $n_{\rm HB}$ gradually increases from 220~K, rises sharply between 140 and 200~K, and reaches its maximum value of 2 at 100~K.
Figure~\ref{MDfig01}(d) depict the representative patterns of hydrogen bonds shown by the arrows at 80~K, 130~K, 180~K, and 300~K. 
At 300~K, the hydrogen-bonded molecules are sparse, among which the population of dimers is the largest. 
At the temperature of 180~K, where $n_{\rm HB}$ changes most rapidly, water molecules have one or two hydrogen bonds and together form structures of hexagonal rings and a variety of chains. 
Below $T_\mathrm{c1}\approx 150$~K, the number of hydrogen-bonded hexamers increase rapidly (Fig.~\ref{MDfig01}(c)), forming a regular array of hexamers.
Each hexamer has clockwise (CW) or counterclockwise (CCW) proton ordering as shown by the green crosses and yellow circles, respectively. 
The formation and breaking of hydrogen bonds takes place at high temperatures, but once the regular array of hexamers is established, the hydrogen bonds are stable and no rearrangement occurs. 
The simulated transition from the orientationally disordered state (HT) to the array of hexamers (LT) agrees with the XRD results, indicating that the water molecules in martyite are well described by the monolayer ice model.

Figure~\ref{MDfig01}(b) shows the in-plane radial distribution functions between O atoms $g(r)$ at 80~K, 130~K, 180~K, and 300~K. 
At 300~K (HT), one can see the broad distribution in the range of $r$ up to 4.5~\AA, corresponding to the three nearest neighboring water molecules around a central one.
Each molecule forms transient hydrogen bonds with one or two of its three neighbors without preference.
At 130~K (LT), however, the distribution corresponding to the three neighbors splits into the first peak at 3~\AA\ with a broad tail over long distances and the second peak at 4~\AA. 
Using the geometric definition of a hydrogen bond that the O$\cdots$O distance $d_\mathrm{OO}$ is less than 4.0~\AA\ and the angle $\theta$ formed by the O-H (chemical bond) and O$\cdots$O lines is less than 30$^\circ$), the first peak corresponds to two hydrogen-bonding neighbors and the second to one non-hydrogen-bonding neighbor. 
At this temperature, the intermolecular distance of a hydrogen-bonded pair fluctuates with large amplitudes. 
The structure of water at 130 K is essentially the same as that of 80 K and below (Fig.~\ref{MDfig01}(d)), \textit{i.e.}, the triangular arrangement of hydrogen-bonded hexamers, each with CW or CCW proton order.

In the H$_2$O hexamer, H$_2$O molecules form the toroidal ring structure.
Here, a head-to-tail configuration of the electric dipole moments ($\mathbf{D}$) form an electric toroidal dipole moment defined by $\mathbf{A} = \sum \mathbf{r} \times \mathbf{D}$ (Fig.~\ref{XRD}(g)).
Our XRD results reveal that the LT phase of martyite is a ferrotoroidal state of H$_2$O driven by the disorder-order transition.
Such ordering is referred to as a ferroaxial or ferro-rotational order, usually driven by the rotational displacements of the coordination polyhedra (e.g. RbFe(MoO$_4$)$_2$, Ca$_5$Ir$_3$O$_{12}$, NiTiO$_3$, and 1T-TaS$_2$)
\cite{doi:10.7566/JPSJ.93.072001, Jin2020,doi:10.7566/JPSJ.90.063702, Hayashida2020, Liu2023}.
Among them, martyite is the simplest case where electric dipoles of polar molecules cooperatively form the toroidal motif, spontaneously breaking the mirror planes parallel to the $c$ axis.
Here, the intermolecular interactions are relatively simple, and the MD simulation can reproduce the toroidal ordering.
In the simulation, the signs of the electric toroidal moments ($\pm \mathbf{A}$) are more or less random since the MD simulation is not long enough at each temperature (10~ns), and therefore the cooling rate is too fast to reach fully equilibrated states. 
However, once such an initial configuration is given by design, the ferrotoroidal order remains at least up to 100~K, and the internal energy is consistently lower than that of the simulated system without ferrotoroidal order (see SM \cite{suppl}). 
Thus, the ground state of the monolayer ice is a ferrotoroidal order, although the simulation gives the random toroidal pattern as a metastable state.

To reveal the fine structure of the ground state, we design the forced ferrotoroidal state (all hexamers are CW) and see how the molecular arrangements change towards lower temperatures (Fig.~\ref{MDfig01}(e)).
Here, type 0, 1, and 2 hexagons represent the H$_2$O hexamers with zero, one, and two longer hydrogen bonds ($d_\mathrm{OO}>3.35$~\AA).
At 120~K, various types of hexamers, including the broken ones shown by the black dots, are distributed randomly.
By decreasing the temperature, the ratio of type-1 hexamers rapidly increases at $T_\mathrm{c2} \approx 100$~K (Fig.~\ref{MDfig01}(c)).
Here, three type-1 hexamers form a H$_2$O octadecamer (dotted triangles in Fig.~\ref{MDfig01}(e)).
At the lowest temperature (LT'), H$_2$O octadecamers alternatively align along one direction, leading to the orthorhombic distortion from the LT phase.

The hexamers' deformations are likely caused by the lattice constants of the honeycomb lattice, which are determined by the framework size. 
Specifically, the distance of 3.36~\AA \ is significantly larger than 2.76~\AA, the ideal hydrogen bond length in ice Ih.
Thus, the H$_2$O hexamer deforms for the sake of five stronger hydrogen bonds at the expense of one weaker one.
Such structural instability is detected by the dielectric and Raman-scattering experiments at $T_\mathrm{c2} \approx$ 30--50~K (Fig.~\ref{dielectric} and Fig.~S7 \cite{suppl}).

The MD simulation suggests that the LT' phase is partially disordered (see the domain boundary shown by the dashed line in Fig.~\ref{MDfig01}(e)).
The energy cost to create a domain boundary is almost zero within our calculation accuracy.
Besides, the distribution of the deformed hexamers depends on the water model and system size (see SM \cite{suppl}).
Therefore, the ground state of the martyite is highly degenerate due to the multiple ways of deforming the hexamers, indicating that the experimental realization of the long-range ordered H$_2$O octadecamers might be challenging.
Our XRD reveals that the atomic displacement parameters of O are elongated toward the intra-hexamer bond direction (see SM \cite{suppl}), suggesting that local distortion from the regular hexamer may already occur at 30~K.
However, the symmetry lowering from trigonal to orthorhombic structure is not detected within the experimental resolution, probably because 30~K is too close to $T_\mathrm{c2}$.
Lower-temperature diffraction experiments are needed to discuss the symmetry of the LT' phase.

In conclusion, we study the quasi-2D water of crystallization in martyite, which shows the hierarchical ordering at low temperatures.
At room temperature (HT), water molecules are dynamically disordered due to the geometrical frustration of the honeycomb lattice.
Here, the rotational and vibrational motions are allowed in the porous framework.
At $T_\mathrm{c1}$, these water molecules cooperatively form H$_2$O hexamers with hydrogen bonds, leading to the ferrotoroidal order in the LT phase.
Once the H$_2$O hexamer is formed, the rotational motions of H$_2$O are forbidden, while the vibrational motions along the hydrogen bonds are still alive.
Finally at $T_\mathrm{c2}$, these vibrational motions of H$_2$O freeze with the deformation of the H$_2$O hexamers in the LT' phase.
After the deformation, the orientation of the type-1 hexamer (location of the long hydrogen bond) appears as a new degree of freedom.
The long hydrogen bonds favor lying close to each other, leading to the formation of the H$_2$O octadecamers.
These successive phase transitions gradually break the original honeycomb-lattice symmetry and lead the system close to the ground state, which is difficult to achieve on the highly degenerate energy landscape.
This degeneracy is the feature of the monolayer ice, behind the polymorphism of H$_2$O.

\section*{Acknowledgements}
We acknowledge K. Homma, S. Morita, N. Kawashima, D. Hamane, M. Akashi, 
J. Nasu, and S. Hayami for the fruitful discussions.
This work was partly supported by JSPS KAKENHI, Grant-in-Aid for Scientific Research (Nos. 22K14010, 23H04861, 24K06944, 24H01644, and 24H01650).
The synchrotron radiation experiments were performed at SPring-8 with the approval of the Japan Synchrotron Radiation Research Institute (JASRI) (Proposal Nos. 2023A1687 and 2024A1709).
The crystal structure figures have been created by using the visualization software VESTA \cite{Momma:ko5060}.


\bibliography{b2}


\clearpage

\setcounter{page}{1}
\title{Supplemental Material of "Ferroaxial order of the monolayer ice in martyite"}

\maketitlesup

\renewcommand{\figurename}{Fig. S}
\setcounter{figure}{0}

\renewcommand{\tablename}{Tab. S}
\setcounter{table}{0}

\section{Methods}

\textit{Sample preparation.}
We grew single- and polycrystals of martyite by hydrothermal synthesis.
We put 100~mg of ZnO (99.0+\%, Fujifilm Wako) powder, 200~mg of V$_2$O$_5$ (99.0+\%, Fujifilm Wako) powder, 200~mg of citric acid, and 50~ml of deionized water in Teflon-lined stainless steel autoclaves, and kept them at 140 $^{\circ}$C for 10--30 days.
Hexagonal transparent single crystals up to the size of 0.5$\times$0.5$\times$0.1 mm$^3$ were obtained.

\textit{Dielectric measurement.}
We measured the dielectric properties of polycrystalline martyite pellets by using an LCR meter (Keysight E4980A).
We used silver paint as an electrode and attached the gold wires on both sides of the pellet.
To avoid water loss from the crystal, we quickly exchanged the sample chamber with He gas and performed the low-temperature experiment down to 4 K.
We note that results were sample dependent, probably because the dielectric properties are sensitive to the water loss and surface adsorbed water.
Nevertheless, the two anomalies at $T_\mathrm{c1}$ and $T_\mathrm{c2}$ are reproduced.

\textit{Single-crystal XRD.}
We performed the x-ray diffraction (XRD) experiments using a single crystal of 80$\times$70$\times$10~$\mu$m$^3$ on BL02B1 at a synchrotron facility SPring-8 in Japan \cite{10.1063/1.3463359}. 
We employed a He-gas-blowing device for controlling the temperature between 30 and 300 K. 
The X-ray wavelength was $\lambda = $ 0.30918 \AA. 
We used a two-dimensional detector CdTe PILATUS to record the diffraction pattern. 
The intensities of Bragg reflections with the interplane distance $d > 0.3$ \AA  
were collected by CrysAlisPro program \cite{CrysAlisPro}. 
Intensities of equivalent reflections were averaged and the structural parameters were refined by using Jana2006 \cite{Jana}.

\textit{Raman scattering.}
We obtained Raman scattering spectra on a single crystal of martyite with the size of 0.5$\times$0.4$\times$0.1~mm$^3$ using a quasi-backscattering geometry.
We used a 532 nm single-frequency laser (Cobolt Samba) with linear polarization for the incident light.
We analyzed the scattered light using a triple monochromator (JASCO NR-1800) equipped with a linear polarizer and a CCD detector (Princeton Instruments Inc. LN/CCD-1100PB).
Temperature dependence above 80~K was conducted using THMS600 (LINKAM SCIENTIFIC) with liquid-N$_2$ under a vacuum environment, and below 150~K using a cryostat with GM refrigerator under He atmosphere.
We measured Raman spectra on the as-grown $c$ surface of martyite with parallel and perpendicular polarization geometries of the scattered light to the incident light.

\textit{MD simulation.}
We carried out molecular dynamics (MD) simulations of two-dimensional water in martyite using GROMACS 2018 software \cite{abraham2015softwarex}. 
The model system consists of $N_w$ water molecules confined between two atomic layers of martyite. 
The atomic positions of martyite other than those of water molecules were fixed during the simulations.	
The number $N_w$ of water molecules was 288 unless otherwise stated. The simulation cell was a rectangular prism under periodic boundary conditions in the two dimensions parallel to the atomic layers. The temperature was controlled by the Nos\'{e}-Hoover thermostat and the volume of the system was fixed.

\section{Single-crystal XRD}
\begin{figure*}[tb]
\centering
\includegraphics[width=16.6cm]{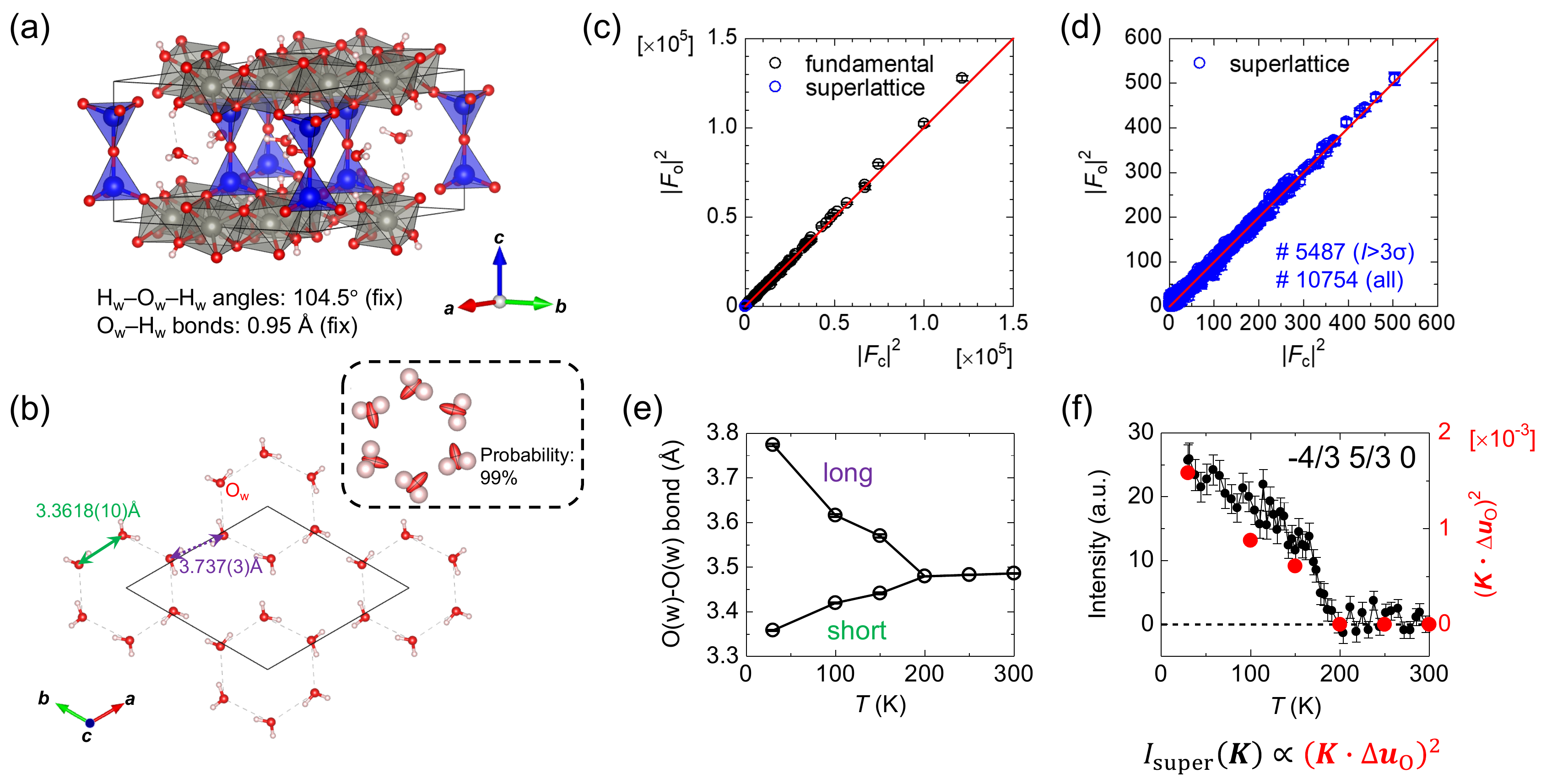}
\caption{\label{fig:xrdanal}
Structural analysis assuming $P\overline{3}$ at 30 K.
(a) Crystal structure of martyite at 30 K. (b) Arrangement of H$_2$O molecules between layers. A H$_2$O hexamer within the black dashed line shows the atomic displacement parameters, indicated by ellipsoids. $|F_\mathrm{c}|^2$--$|F_\mathrm{o}|^2$ plot of the structural analysis at 30 K showing (c) fundamental and superlattice reflections and (d) only superlattice reflections. (e) Temperature dependence of the O$_\mathrm{w}$--O$_\mathrm{w}$ bond distances in the H$_2$O layer. (f) Temperature dependence of the superlattice peak intensity of -4/3 5/3 0. The black dots represent the observed intensity, while the red dots represent the calculations that only consider the O$_\mathrm{w}$ displacement from the $P\overline{3}m1$ structure in the HT phase.
} 
\end{figure*}

The results of the structural analysis of martyite are summarized in Tables S~I--S~IV and Figures~S\ref{fig:xrdanal}--S\ref{fig:origin}. 
Figures~S\ref{fig:xrdanal}(a) and S\ref{fig:xrdanal}(b) show the crystal structure at 30 K (LT) with the space group $P\overline{3}$. 
In the structural refinement, we fixed the O$_\mathrm{w}$--H$_\mathrm{w}$ bond length and H$_\mathrm{w}$-O$_\mathrm{w}$-H$_\mathrm{w}$ bond angle in H$_2$O molecules at 0.95 \AA \ and \ang{104.5}, respectively. 
Figure~S\ref{fig:xrdanal}(c) shows the $|F_\mathrm{c}|^2$--$|F_\mathrm{o}|^2$ plot resulting from the structural analysis, and Fig.~S\ref{fig:xrdanal}(d) displays only the weak superlattice reflections. 
Here, $F_\mathrm{c}$ and $F_\mathrm{o}$ indicate the calculated and observed crystal structure factor, respectively. 
The $|F_\mathrm{c}|^2$ and $|F_\mathrm{o}|^2$ of each diffraction reflection show high linearity, indicating good agreement between the experimental and calculated intensities.

Figure~S\ref{fig:xrdanal}(e) shows the temperature dependence of the O$_\mathrm{w}$--O$_\mathrm{w}$ bond distances in the H$_2$O layer. 
Figure~S\ref{fig:xrdanal}(f) shows the temperature dependence of the superlattice peak intensity. 
The black dots indicate the observed intensity of -4/3 5/3 0, while the red dots represent the calculated intensity of -4/3 5/3 0, considering only the O$_\mathrm{w}$ displacement $\Delta \bm{u}_\mathrm{O}$ from the $P\overline{3}m1$ structure. 
The superlattice peak intensity scales primarily with the displacement of O atoms within the H$_2$O molecule.

\begin{figure*}[tb]
\centering
\includegraphics[width=15.9cm]{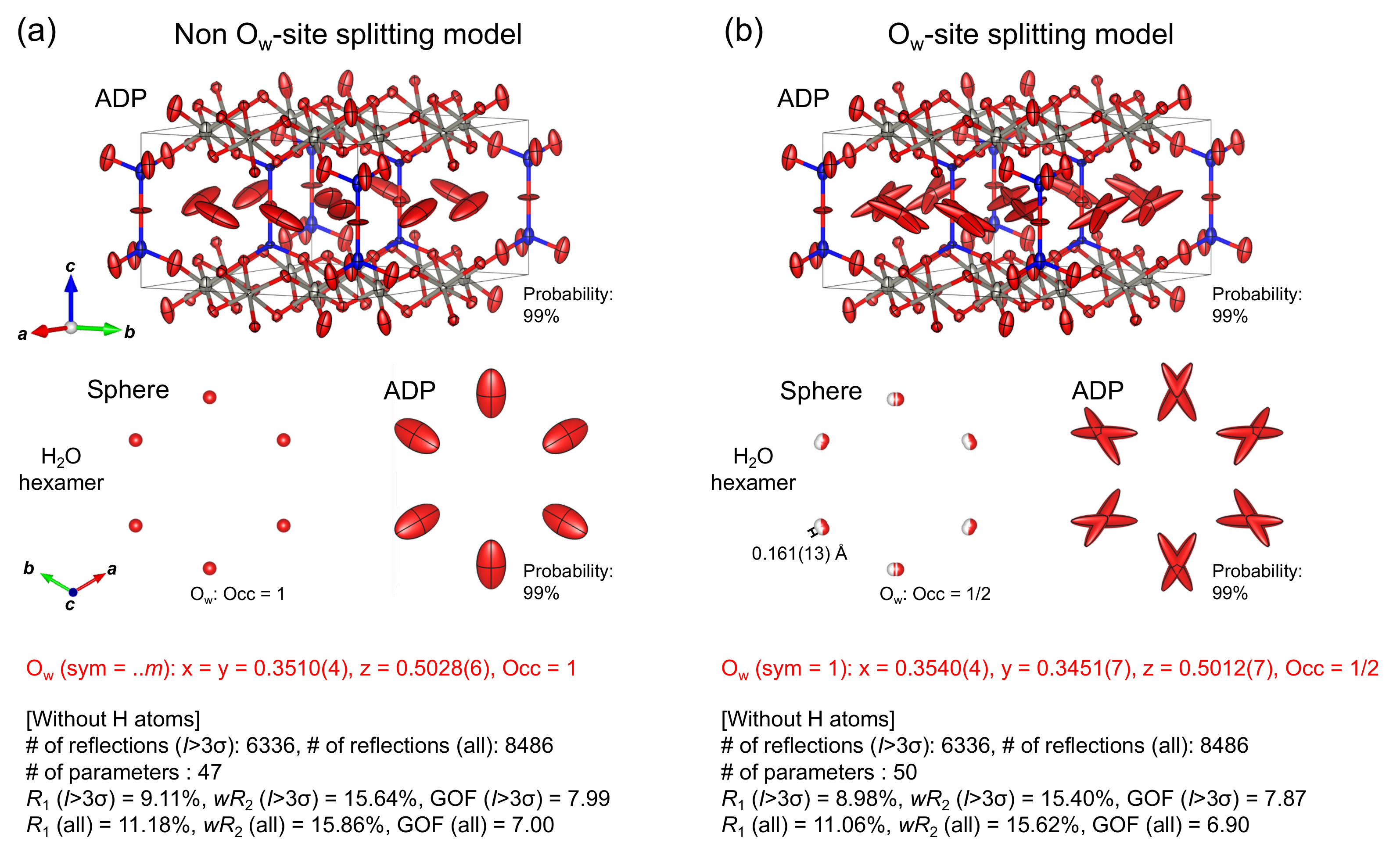}
\caption{\label{fig:splitmodel}
Structural analysis assuming $P\overline{3}1m$ at 30 K.
Results of the structural analysis at 30 K, assuming the $P\overline{3}1m$ space group. 
Figures (a) and (b) show the analysis results before and after incorporating the O$_\mathrm{w}$-site split model, respectively. 
Atomic positions are represented as spheres, while atomic displacement parameters (ADPs) are depicted as ellipsoids.
} 
\end{figure*}

If the toroidal arrangement of water molecules is randomly distributed between the nearest hexamers, the structure can be analyzed using a model that assumes the disorder of water molecules within the $P\overline{3}1m$ space group. 
Specifically, this can be represented by a split model for the O$_\mathrm{w}$ atoms that constitute the water molecules. 
We performed structural analysis at 30 K, assuming the $P\overline{3}1m$ space group. 
The analysis results before and after incorporating the O$_\mathrm{w}$-site split model are shown in Fig. S\ref{fig:splitmodel}(a) and S\ref{fig:splitmodel}(b), respectively. 
Here, the occupancy of the O$_\mathrm{w}$ site is fixed at 1/2. 
Note that when analyzing a model including hydrogen atoms, the parameters did not converge. 
Therefore, we present analyses that exclude hydrogen atoms. 
The split width corresponding to the O$_\mathrm{w}$-O$_\mathrm{w}$ distance is 0.161(13) \AA. 
When the O$_\mathrm{w}$-site split model is applied, the $R$ values of the analysis improve; however, it remains significantly worse than that obtained under the assumption of the $P\overline{3}$ space group (Table S II). 
These results suggest that, in the low-temperature phase, the toroidal order of water molecules is not random but exhibits a ferroic arrangement within the plane.

\begin{figure*}[tb]
\centering
\includegraphics[width=14.9cm]{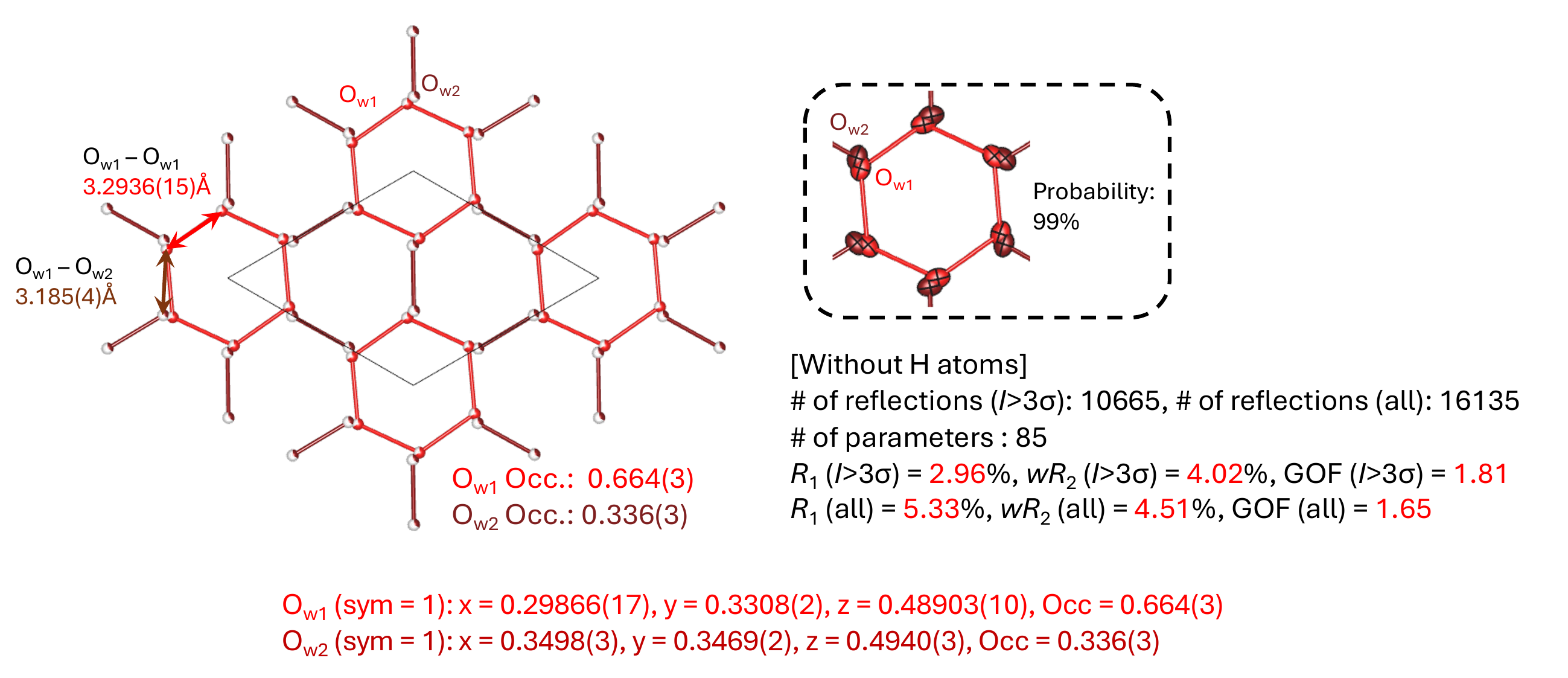}
\caption{\label{fig:split}
O site split model analysis assuming $P\overline{3}$ at 30 K.
(a) Results of the O$_\mathrm{w}$ split model analysis at 30 K. 
A H$_2$O hexamer within the black dashed line shows the atomic displacement parameters, indicated by ellipsoids. 
} 
\end{figure*}

\begin{figure*}[tb]
\centering
\includegraphics[width=16.6cm]{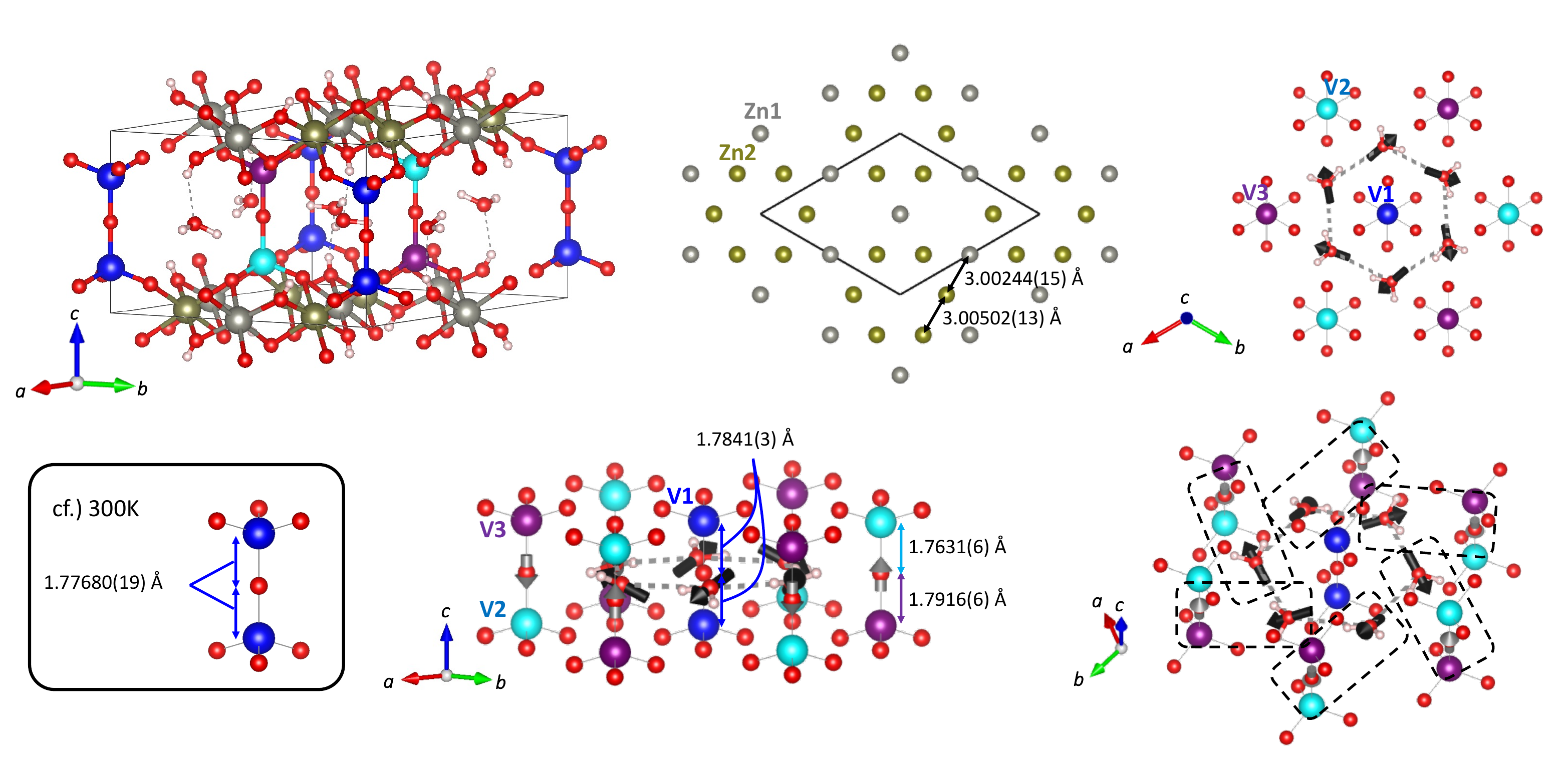}
\caption{\label{fig:origin}
Crystal structure of martyite at 30 K. 
The Zn layer changes little during the phase transition, while in the V$_2$O$_7$ pillar, the O atom between the V atoms displaces by up to 1~\% along the $c$-axis direction. 
The electric dipole moment due to the displacement of the O atoms in the V$_2$O$_7$ pillars interacts with the dipole moment of H$_2$O molecules, resulting in a ferrotoroidal order not only in-plane but also in the stacking direction.
} 
\end{figure*}

Assuming the $P\overline{3}$ space group, the atomic displacement parameter of O$_\mathrm{w}$ in the H$_2$O hexamer at 30 K extends along the bonding direction, as shown in the inset of Fig.~S\ref{fig:xrdanal}(b).
Therefore, we performed the structural analysis using a site splitting model for O$_\mathrm{w}$. 
Figure S\ref{fig:split} shows the results of the splitting model analysis at 30 K, where the H$_\mathrm{w}$ atoms of the H$_2$O molecule have been removed. 
All structural parameters were converged, with the occupancy ratio of the split oxygen atoms being approximately O$_\mathrm{w1}$:O$_\mathrm{w2}$ = 2:1. 
Consequently, shorter O$_\mathrm{w1}$–O$_\mathrm{w1}$ and O$_\mathrm{w1}$–O$_\mathrm{w2}$ bond distances were obtained compared to the non-split model, which may be related to the deformation of water hexamers observed in the MD simulations.

Considering the dipole-dipole interactions of water molecules between the layers, an anti-ferroic toroidal arrangement along the interlayer direction is more stable than the ferroic one observed in the experiment (see also Fig.~S\ref{fig:energy}(b)).
We examine the relationship between the framework structure--composed of ZnO$_4$(OH)$_2$ octahedra and V$_2$O$_7$ pillars--and the interlayer water molecules. 
Figure~S\ref{fig:origin} presents the crystal structure at 30 K, highlighting the ZnO$_4$(OH)$_2$ layers and V$_2$O$_7$ pillars. 
The structure contains two Zn sites and three V sites with distinct symmetries. In the V(1)$_2$O$_7$ pillar, the distance between the V atoms above and below the central O is uniform. 
However, in the V(2)V(3)O$_7$ pillar, the V-O distances above and below the central O differ, indicating an electric dipole moment along the $c$-axis, represented by grey vectors. 
This dipole moment, originating from the V(2)V(3)O$_7$ pillar, shows a ferroic arrangement along the $c$-axis and appears energetically stable. 
Notably, the $c$-axis component of the electric dipole moment from the adjacent water molecules opposes that of the V$_2$O$_7$ pillar. 
This suggests that the ferroic deformation of the V(2)V(3)O$_7$ induces the local dipole moment in the framework and stabilizes the ferroic toroidal arrangement of the water molecules' dipole moments between the layers.

\begin{table}[tb]
    \caption{\label{tb:300K} Summary of crystallographic data of martyite at 300~K (HT).}
    \centering
    \begin{ruledtabular}
    \begin{tabular}{l l}
        Wavelength (\AA) & 0.30918  \\
        Crystal dimension ($\mu$m$^3$) & 80$\times$70$\times$10  \\
        Space group & $P\overline{3}m1$  \\
        $a$ (\AA) & 6.0325(3) \\
        $c$ (\AA) & 7.1667(5)  \\
        Chemical formula & Zn$_3$V$_2$O$_7$(OH)$_2$$\cdot$2H$_2$O  \\
        $Z$ & 1  \\
        $F(000)$ & 230  \\
        ($\sin \theta$/$\lambda$)$_\mathrm{max}$ (\AA$^{-1}$) & 1.67  \\
        $N_\mathrm{total}$ & 19686  \\
        $N_\mathrm{unique}$ (all) & 3017  \\
        $N_\mathrm{unique}$ ($I>3\sigma$) & 2089  \\
        Average redundancy & 6.525  \\
        Completeness (\%) & 95.14  \\
        $N_\mathrm{parameters}$ & 25  \\
        $R_1$ ($I>3\sigma$ / all) (\%) & 2.78 / 4.82 \\
        $wR_2$ ($I>3\sigma$ / all) (\%) & 3.89 / 4.52 \\
        GOF ($I>3\sigma$ / all) & 1.58 / 1.50 \\
        CCDC & 2425731  \\
    \end{tabular}
    \end{ruledtabular}
\vspace{1cm}
    \caption{\label{tb:30K} Summary of crystallographic data of martyite at 30~K (LT).}
    \centering
    \begin{ruledtabular}
    \begin{tabular}{l l}
        Wavelength (\AA) & 0.30918  \\
        Crystal dimension ($\mu$m$^3$) & 80$\times$70$\times$10  \\
        Space group & $P\overline{3}$  \\
        $a$ (\AA) & 10.41250(10)  \\
        $c$ (\AA) & 7.14370(10) \\
        Chemical formula & Zn$_3$V$_2$O$_7$(OH)$_2$$\cdot$2H$_2$O  \\
        $Z$ & 3  \\
        $F(000)$ & 690  \\
        ($\sin \theta$/$\lambda$)$_\mathrm{max}$ (\AA$^{-1}$) & 1.67  \\
        $N_\mathrm{total}$ & 117979  \\
        $N_\mathrm{unique}$ (all) & 16135  \\
        $N_\mathrm{unique}$ ($I>3\sigma$) & 10665  \\
        Average redundancy & 7.312  \\
        Completeness (\%) & 94.38  \\
        $N_\mathrm{parameters}$ & 85  \\
        $R_1$ ($I>3\sigma$ / all) (\%) & 3.19 / 5.54 \\
        $wR_2$ ($I>3\sigma$ / all) (\%) & 4.23 / 4.69 \\
        GOF ($I>3\sigma$ / all) & 1.91 / 1.72 \\
        Flack parameters ($a,b,c : b,a,c$) & 0.4992(18):0.5008  \\
        CCDC & 2425733  \\
    \end{tabular}
    \end{ruledtabular}
\end{table}

\begin{turnpage}
\begin{table*}[tbh]
    \caption{\label{tb:300K_parameter} Structural parameters of martyite at 300 K (HT). The space group is $P\overline{3}m1$ (No. 164) and $a = 6.0325(3)$~\AA, $c = 7.1667(5)$~\AA.}
    \centering
    \begin{ruledtabular}
    \begin{tabular}{lccrrrrrrrrrr}        
Atom & Wyckoff & Symmetry & $x$ & $y$ & $z$& Occ. & $U_{11}$ (\AA$^2$) & $U_{22}$ (\AA$^2$) & $U_{33}$ (\AA$^2$) & $U_{12}$ (\AA$^2$) & $U_{13}$ (\AA$^2$) & $U_{23}$ (\AA$^2$)  \\
& position & &&&&& & & & & & \\ \hline
Zn & $3e$ & .2/m. & 1/2 & 0 & 0 & 1 & 0.01484(4) & 0.00968(3) & 0.02778(7) & 0.004840(17) & -0.002628(16) & -0.00526(3)  \\
V & $2c$ & 3m. & 0 & 0 & 0.747924(19) & 1 & 0.00935(3) & 0.00935(3) & 0.00759(3) & 0.004673(14) & 0 & 0  \\
O1 & $2d$ & 3m. & 1/3 & 2/3 & 0.87724(10) & 1 & 0.01010(11) & 0.01010(11) & 0.01140(18) & 0.00505(5) & 0 & 0  \\
O2 & $1b$ & -3m. & 0 & 0 & 1/2 & 1 & 0.0249(4) & 0.0249(4) & 0.0102(3) & 0.01247(19) & 0 & 0  \\
O3 & $6i$ & .m. & 0.30855(10) & 0.15427(5)	& 0.82608(8) & 1 & 0.01354(13) & 0.01379(10) & 0.01584(15) & 0.00677(7) & -0.00234(11) & -0.00117(6)\\
O4w & $2d$ & 3m. & 2/3 & 1/3 & 0.4890(3) & 1 & 0.0911(17) & 0.0911(17) & 0.0169(6) & 0.0456(8) & 0 & 0  \\
H1 & $2d$ & 3m. & 1/3 & 2/3 & 0.74469(10) & 1 & ~ & ~ & ~ & ~ & ~ &   \\
H2w & $6i$ & .m. & 0.5784(15) & 0.157(3) & 0.521(9) & 2/3 (fix) & ~ & ~ & ~ & ~ & ~ & ~ \\
    \end{tabular} 
    \end{ruledtabular}
\vspace{1cm}
    \caption{\label{tb:300K_parameter} Structural parameters of martyite at 30 K (LT). The space group is $P\overline{3}$ (No. 147) and $a = 10.41250(10)$~\AA, $c = 7.14370(10)$~\AA.}
    \centering
    \begin{ruledtabular}
    \begin{tabular}{lccrrrrrrrrr}
Atom & Wyckoff & Symmetry & $x$ & $y$ & $z$ & $U_{11}$ (\AA$^2$) & $U_{22}$ (\AA$^2$) & $U_{33}$ (\AA$^2$) & $U_{12}$ (\AA$^2$) & $U_{13}$ (\AA$^2$) & $U_{23}$ (\AA$^2$)  \\
& position & &&&& & & & & & \\ \hline
Zn1 & $3e$ & -1 & 1/2 & 1/2 & 0 & 0.00474(5) & 0.00581(6) & 0.00876(6) & 0.00242(3) & 0.00155(4) & 0.00175(4) \\
Zn2 & $6g$ & 1 & 0.333132(13) & 0.166061(9) & 0.005430(10) & 0.00457(2) & 0.00298(3) & 0.005506(17) & 0.00152(3) & 0.00073(4) & -0.00013(2) \\
V1 & $2c$ & 3.. & 0 & 0 & 0.25025(4) & 0.00405(4) & 0.00405(4) & 0.00485(6) & 0.002026(18) & 0 & 0  \\
V2 & $2d$ & 3.. & 2/3 & 1/3 & 0.25557(2) & 0.00377(3) & 0.00377(3) & 0.00274(3) & 0.001885(12) & 0 & 0  \\
V3 & $2d$ & 3.. & 2/3 & 1/3 & 0.75317(2) & 0.00357(2) & 0.00357(2) & 0.00220(3) & 0.001784(12) & 0 & 0  \\
O1 & $1b$ & -3.. & 0 & 0 & 1/2 & 0.0110(4) & 0.0110(4) & 0.0033(4) & 0.00552(18) & 0 & 0  \\
O2 & $2d$ & 3.. & 2/3 & 1/3 & 0.50240(7) & 0.00756(16) & 0.00756(16) & 0.0050(2) & 0.00378(8) & 0 & 0  \\
O3 & $6g$ & 1 & 0.51334(7) & 0.33407(14) & 0.83262(6) & 0.00166(13) & 0.00889(14) & 0.00631(9) & 0.00116(14) & 0.00027(10) & -0.00129(18) \\
O4 & $6g$ & 1 & 0.15510(10) & -0.00107(12) & 0.17192(12) & 0.0108(2) & 0.00339(14) & 0.0092(2) & 0.00068(18) & -0.00001(18) & -0.00157(17) \\
O5 & $6g$ & 1 & 0.82202(8) & 0.33391(14) & 0.17835(6) & 0.00575(19) & 0.00763(13) & 0.00676(10) & 0.00403(18) & 0.00033(12) & -0.0019(2) \\
O6 & $6g$ & 1 & 0.33317(7) & 0.33491(7) & 0.12567(4) & 0.00484(10) & 0.00441(10) & 0.00539(6) & 0.00124(14) & 0.00163(13) & 0.00136(13) \\
O7w & $6g$ & 1 & 0.30981(14) & 0.33398(12) & 0.49011(9) & 0.0809(8) & 0.0185(3) & 0.00842(14) & 0.0243(5) & 0.0028(4) & 0.0011(3) \\
H1w & $6g$ & 1 & 0.3844(13) & 0.4228(12) & 0.546(3) & ~ & ~ & ~ & ~ & ~ &   \\
H2w & $6g$ & 1 & 0.344(2) & 0.2654(17) & 0.506(2) & ~ & ~ & ~ & ~ & ~ &   \\
H3 & $6g$ & 1 & 0.2942(17) & 0.2974(17) & 0.2462(11) & ~ & ~ & ~ & ~ & ~ & ~ \\
    \end{tabular} 
    \end{ruledtabular}
\end{table*}
\end{turnpage}

\clearpage

\section{Energy scale of the intermolecular interactions}

\begin{figure}[tb]
\centering
\includegraphics[width=8.3cm]{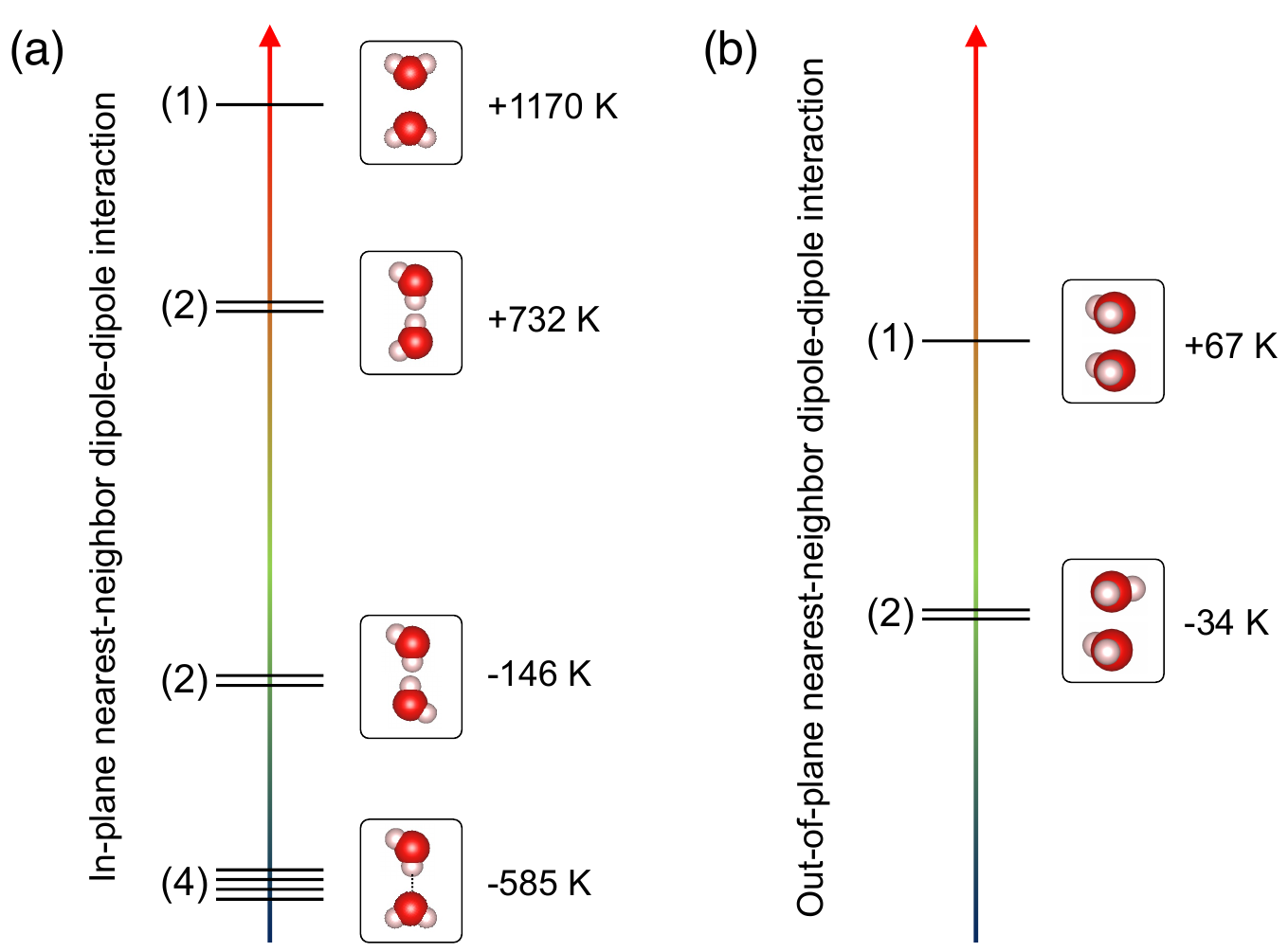}
\caption{\label{fig:energy}
Dipole-dipole interaction energies between water molecules.
(a) In-plane nearest-neighbor interactions. (b) Out-of-plane nearest-neighbor interactions.
The numbers in the brackets indicate the degeneracy.
} 
\end{figure}

The crystal structure of martyite suggests that the water of crystallization is a quasi-2D system where each water layer is separated by the ZnO$_4$(OH)$_2$ block layers.
Here, each H$_2$O molecule can be hydrogen-bonded only with the in-plane nearest neighbors.
The second important interaction is the electric dipole-dipole interaction.
In this section, we evaluate the dipole-dipole interaction energy scale in martyite.
For simplicity, we neglect the electric polarization of the framework of martyite and treat the intermolecular medium as a vacuum.
The intermolecular distances are fixed at the experimental values determined by the single-crystal XRD at 300~K.

Figure S\ref{fig:energy}(a) shows the dipole-dipole interaction energies of the in-plane nearest-neighbor pair, $E_\mathrm{iNN}$.
Here, we assume that H$_2$O molecules at A and B sites can take only three orientations as suggested by the XRD at 300~K.
The out-of-plane displacement of H$_2$O, which is suggested from the low-temperature XRD, is not taken into account.
The interaction energy is estimated as $E_\mathrm{iNN}=$ $+1170$~K, $+732$~K, $-146$~K, $-585$~K, from top to bottom molecular geometries.
Considering the energy scale of hydrogen bonds from $-2000$~K to $-3000$~K, the bottom geometry is further stabilized.
Therefore, water molecules tend to increase the number of hydrogen bonds towards low temperatures.
When the hydrogen-bonded H$_2$O hexamers are formed, two types of hexamers appear with CW or CCW electric toroidal dipole moment.
The inter-hexamer interaction is determined by the molecular geometry between two hexamers.
This energy diagram suggests that the second geometry from the bottom ($-146$~K) is much stabler than the third one ($+732$~K).
Indeed, the XRD results suggest that the ferrotoroidal order of H$_2$O is realized in the LT phase, where the inter-hexamer molecular geometry is of this type.

Second, we estimate the dipole-dipole interaction energy of the out-of-plane nearest-neighbor pair, $E_\mathrm{oNN}$.
Similarly, we assume that H$_2$O molecules can take only three orientations.
The interaction energy is estimated as $E_\mathrm{oNN}=$ $+67$~K (top) and $-34$~K (bottom).
This energy scale is much smaller than the in-plane one, with the typical ratio of 50~K/2500~K = 2~\%, including the hydrogen bonds.
Therefore, as indicated by the crystal structure, martyite can be considered a quasi-2D system.
We should note, however, that the out-of-plane interactions can be mediated via the 3D framework deformation as Fig.~S\ref{fig:origin}.
Nevertheless, the agreement between the experiments and the MD simulations on the monolayer ice suggests that the expression "monolayer ice in martyite" is not just an exaggeration, but captures the features of this mineral.

\section{Raman scattering}

\begin{figure*}[tbh]
\centering
\includegraphics[width=17cm]{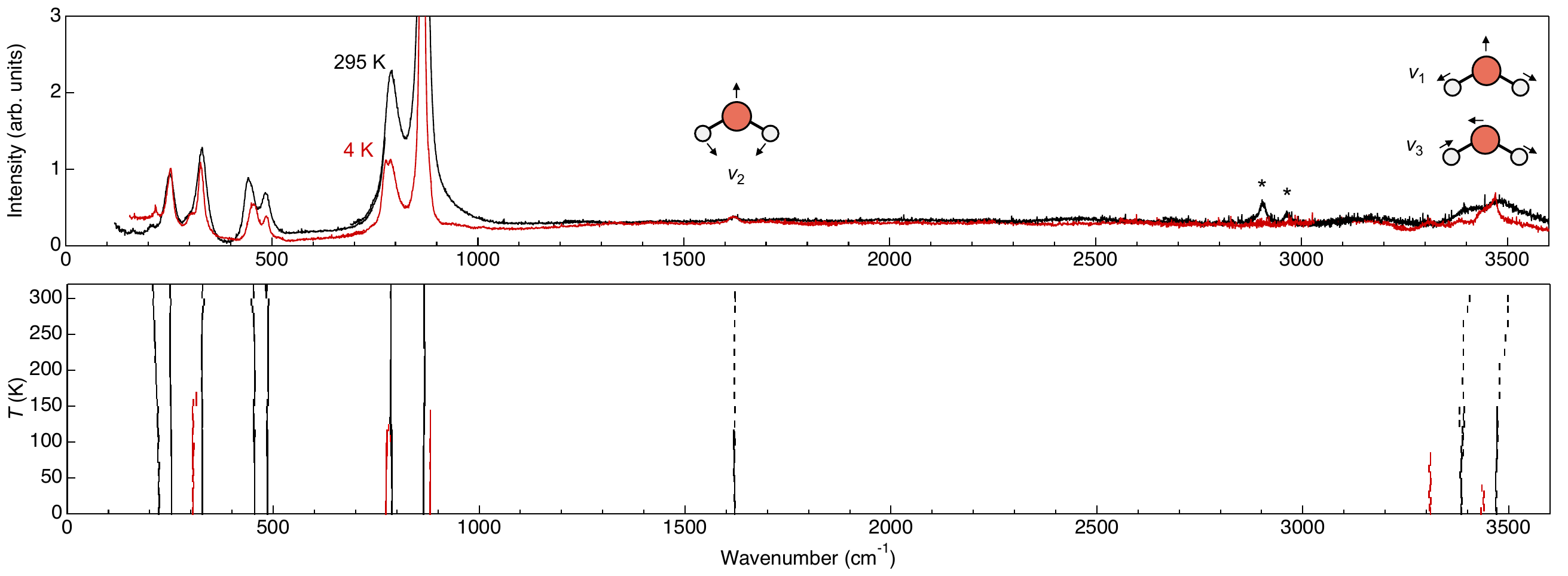}
\caption{\label{fig:raman}
Raman spectra of martyite.
The upper panel shows the Raman scattering spectra at 295~K and 4~K.
The schematic vibrational modes of H$_2$O ($v_1$, $v_2$, and $v_3$) are also shown.
The asterisks indicate the peaks due to the grease for sample mounting.
The lower panel shows the assigned peak energies at each temperature.
The red symbols indicate the peaks observed only at lower temperatures.
} 
\end{figure*}

The upper panel of Fig. S\ref{fig:raman} shows the Raman scattering spectra of martyite at 295~K and 4~K.
The Raman peaks can be assigned as the framework or the H$_2$O molecules' origins.
Based on the numerical simulation, we conclude that the lower-energy peaks from 240 to 900~cm$^{-1}$ originate from the framework vibration.
H$_2$O molecule has three vibrational modes, O-H symmetric streching ($v_1$), H-O-H bending ($v_2$), and O-H asymmetric streching ($v_3$) modes \cite{Scherer1977}.
We assign that the relatively weak peak at 1620~cm$^{-1}$ is $v_2$ mode and the band from 3300 to 3600~cm$^{-1}$ corresponds to the superposition of $v_1$ and $v_3$ modes.
Although the O-H vibration (the $A_g$ representation mode under $\overline{3}m$ point group) of the martyite framework is also Raman active, it was not observed probably due to weak intensity.
The asterisks indicate the artefacts due to the grease for sample mounting.

\begin{figure}[tbh]
\centering
\includegraphics[width=8.2cm]{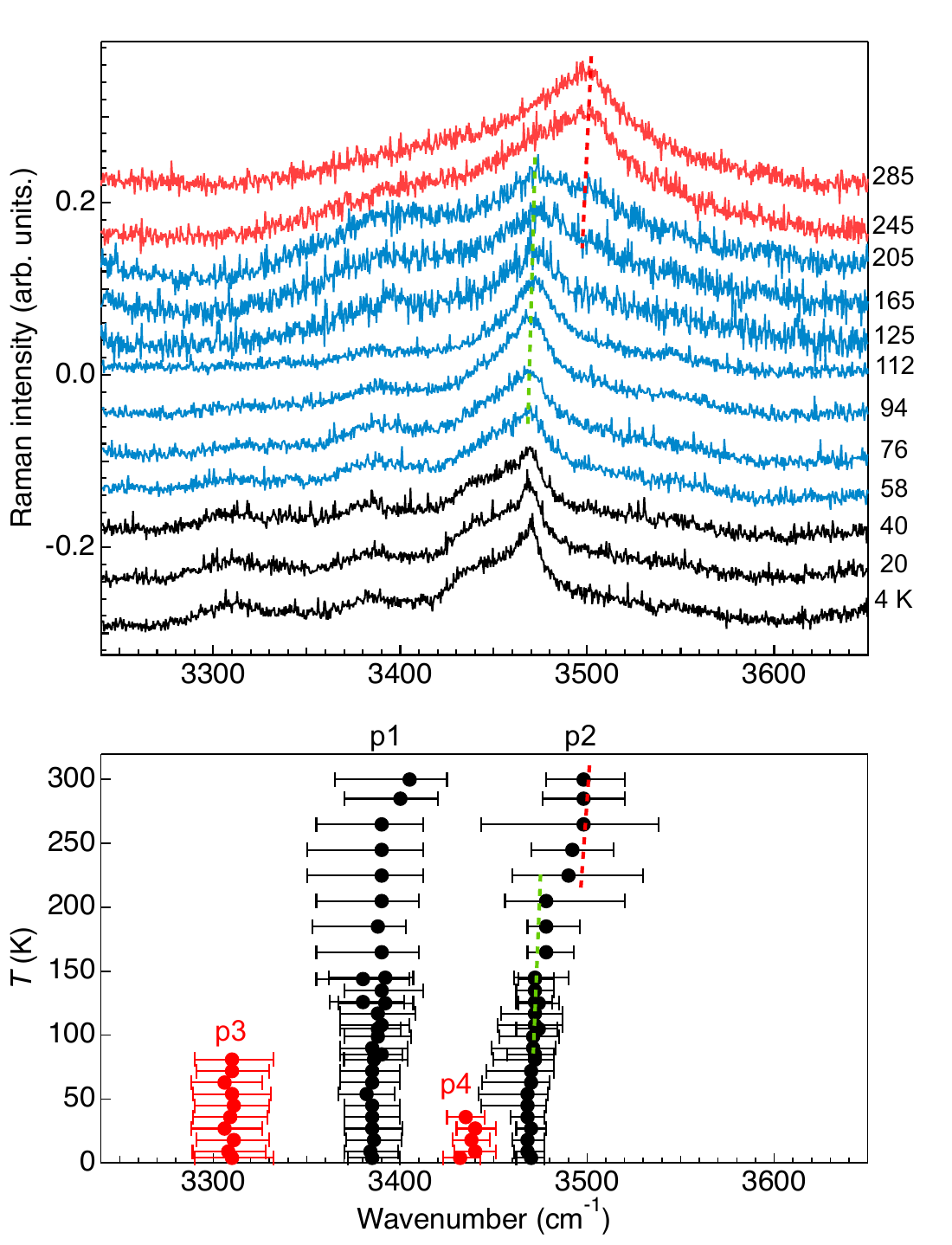}
\caption{\label{fig:raman3400}
Stretching modes of the Raman spectra.
The upper panel shows the enlarged Raman scattering spectra around $v_1$ and $v_3$ from 285~K to 4~K.
The lower panel shows the assigned peak energies (p1--p4) at each temperature.
The red and green dashed lines are shown for the guide.
} 
\end{figure}

The lower panel of Fig.~S\ref{fig:raman} shows the peak energies at each temperature.
Red symbols indicate the peak appearing at lower temperatures.
The peaks at 305~cm$^{-1}$, 774~cm$^{-1}$, 882~cm$^{-1}$ appear at around 150~K, where the ferrotoroidal ordering of H$_2$O occurs at $T_\mathrm{c1}$ (LT).
These new peaks indicate the symmetry breaking of the framework as suggested by the XRD experiment.
Among the vibrational modes of H$_2$O, the $v_2$ peak does not show any peak shift or splitting.
In contrast, $v_1$ and $v_3$ modes around 3400~cm$^{-1}$ show complicated peak splitting patterns toward low temperatures.

Figure~S\ref{fig:raman3400} shows the enlarged Raman spectra of martyite from 285~K to 4~K.
The upper panel presents the Raman spectra with red (HT), blue (LT), and black (LT') curves.
HT phase show broad 2 peaks at 3500~cm$^{-1}$ and at 3390~cm$^{-1}$, probably corresponding to $v_1$ and $v_3$.
Because of the light reflection at the bottom surface of the transparent crystal, we could not see a clear difference in the parallel and perpendicular polarization of the scattered light.
Thus, we could not conclude which peak corresponds to $v_1$ and $v_3$ modes.
Since the energy relationship ($v_1 < v_3$ or $v_3 < v_1$) depends on the potential shape around H$_2$O, it is also challenging to estimate from the numerical calculation.
Therefore, we keep this assignment for a future issue.

In the LT phase, the peak at 3500~cm$^{-1}$ (red dashed line) disappears, and a new peak at 3480~cm$^{-1}$ (green dashed line) appears.
The S/N ratio changed from 125~K to 112~K because we used a different cryostat for the low-temperature experiments.
Under the determined $P\overline{3}$ crystal structure, there are 4 Raman active modes for $v_1$ and $v_3$ vibrations, \textit{i.e.} $2A_g+2E_g$. 
However, the structure contains only one site for H$_2$O molecules. 
Therefore, the result of the two peaks is reasonable. 
They are assigned to $v_1$ or $v_3$ mode of the H$_2$O molecules. 
The energy difference between $A_g$ and $E_g$ mode, which is distinguished by the phase factor between adjacent H$_2$O molecules, is small and is obscured by their line widths.

In the LT' phase (even from higher temperatures around 90~K), new peaks appear at 3310~cm$^{-1}$ and 3440~cm$^{-1}$.
At least, 4 peaks [3390~cm$^{-1}$ (p1), 3480~cm$^{-1}$ (p2), 3310~cm$^{-1}$ (p3), and 3440~cm$^{-1}$ (p4)] are observed at 4~K.
If we assign the 4 modes to $2A_g+2E_g$ in the $P\overline{3}$ structure, the new modes must appear as the modes are splitting. 
However, the p3 is a new peak and does not originate from the p1 peak. 
Therefore, this result suggests that there are at least two H$_2$O sites, and indicates the symmetry lowering probably due to the deformation of the H$_2$O hexamer below $T_\mathrm{c2}$ as suggested from the MD simulation.

\clearpage
\section{MD simulation}

\subsection{Simulation conditions}

The lateral dimensions of the simulation cell for $N_w=288$ are 7.236~nm$\times$ 6.264~nm.
The water model employed in the present study is primarily TIP4P/ICE \cite{10.1063/1.1931662} and, alternatively, TIP4P/2005 \cite{10.1063/1.2121687} is used for comparison. The water-martyite interaction is described by the sum of the pair potentials $\phi_{ij}(r)$ between an interaction site $i$ of the water model and an atom $j$ in martyite:  
$$
\phi_{ij}(r)=4\epsilon_{ij}\left[\left(\frac{\sigma_{ij}}{r}\right)^{12}-\left(\frac{\sigma_{ij}}{r}\right)^{6}\right]
+ \frac{Z_iZ_j e^2}{4\pi \varepsilon_0 r},
$$
where the Lennard-Jones (LJ) size and energy parameters $\sigma_{ij}$  and $\epsilon_{ij}$ 
are given by the geometric averages.
The LJ parameters and charges for the atoms in martyite are listed in Table S\ref{tb:lj}.

\begin{table}[b]
\caption{\label{tb:lj} Lennard-Jones parameters and charges for the atoms of the framework of martyite.}
\begin{ruledtabular}
\begin{tabular}{rllr}
\textrm{Atom}&
$\epsilon$ [kJ/mol] &
$\sigma$ [nm] &
\textrm{$Z$}\\
\colrule
H\footnote{Those of the OH groups facing the layer of water
molecules.}  & 0 & 0 & 1\\
O$^\text{a}$ & 0.7749 & 0.3159 & $-1$\\
H   & 0 & 0 & 0\\
O   & 0.7749 & 0.3159 & 0\\
V  & 0.1338 & 0.2630 & 0\\
Zn & 0.06241 & 0.2486 & 0\\
\end{tabular}
\end{ruledtabular}
\end{table}

Water molecules were first arranged at the honeycomb lattice sites with arbitrary molecular orientations, equilibrated for 1~ns at 5~K with the positional constraints on the water's oxygen atoms, and then further equilibrated for 10 ns at 300 K without any positional constraints. 
At 300 K and below, without any positional constraints, the water molecules oscillate and rotate around each lattice point, but do not migrate to other lattice points or interstitials.
The temperature $T$ was decreased from 300 K to 1 K in steps of 10 K for $T \ge 10$~K and in steps of 1 K for $T < 10$~K. 
The simulated time at each temperature was 10 ns for $T\ge 10$~K, and 3 ns for $T < 10$~K,  the first 1 ns being used for equilibration and the rest for data sampling. 
The MD time step was 1 fs. The cutoff distance for the LJ potentials was 1.0 nm. The Coulomb potentials were treated using the particle mesh Ewald method, with the real-space cutoff distance being the same as the cutoff distance for the LJ potentials. For the cooling process to study structural changes, 5 sets of MD simulations were performed starting from independent initial conditions, and the numerical results are the averages over the 5 sets. 
To examine the effects of system size on the ferrotoroidal ice structure, one set of MD simulations was performed for the large system ($N_w=864$, the lateral dimensions of 10.854~nm~$\times$~12.528~nm, TIP4P/ICE model, $T \le $ 170 K).

\subsection{Movie of simulated water molecules in martyite}
The movie of molecular dynamics of water molecules in martyite is available online. 
The temperature is lowered from 300 K to 10 K in steps of 10 K. 
One second in the movie corresponds to 1.2~ps.
In the movie, one can observe the transition at $T_\text{c1}\approx 190 $~K from a state in which water molecules vibrate vigorously and rotate randomly at around honeycomb lattice sites to a state in which they form six-membered rings through hydrogen bonding, followed by the transition at $T_\text{c2}$ from a state in which the six-membered rings dynamically change their shapes to a state in which the six-membered rings are fixed in the distorted shapes.

\subsection{Comparison of MD simulations with the water models TIP4P/ICE and TIP4P/2005}

\begin{figure}[tb]
\centering
\includegraphics[width=8.2cm]{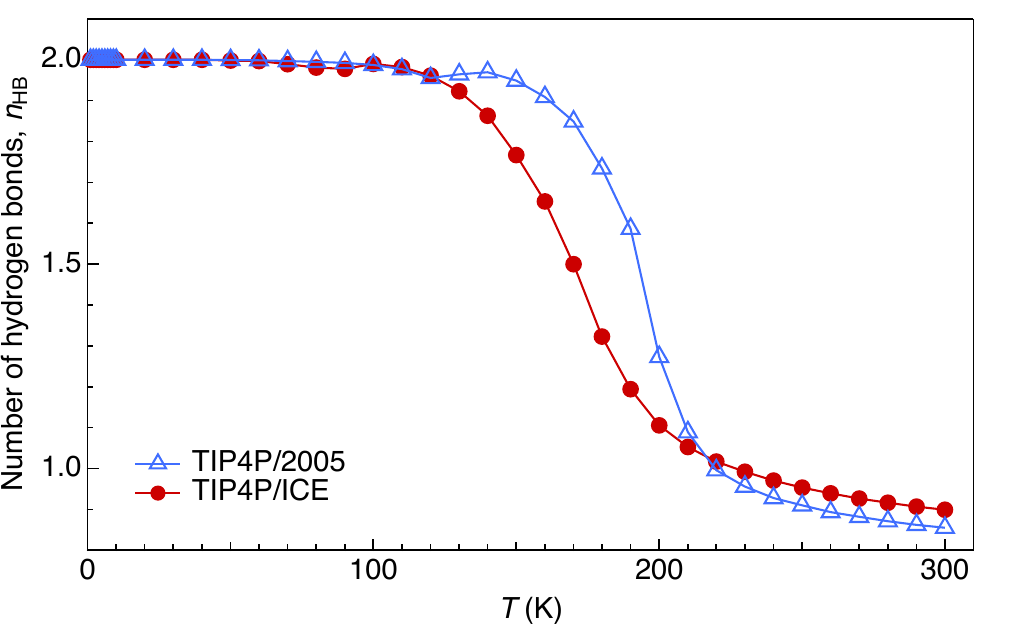}
\caption{\label{fig:n_hb}
Number of hydrogen bonds per water molecule in martyite as a function of temperature.
The water models are TIP4P/2005 and TIP4P/ICE. 
} 
\end{figure}

\begin{figure}[tb]
\centering
\includegraphics[width=8.2cm]{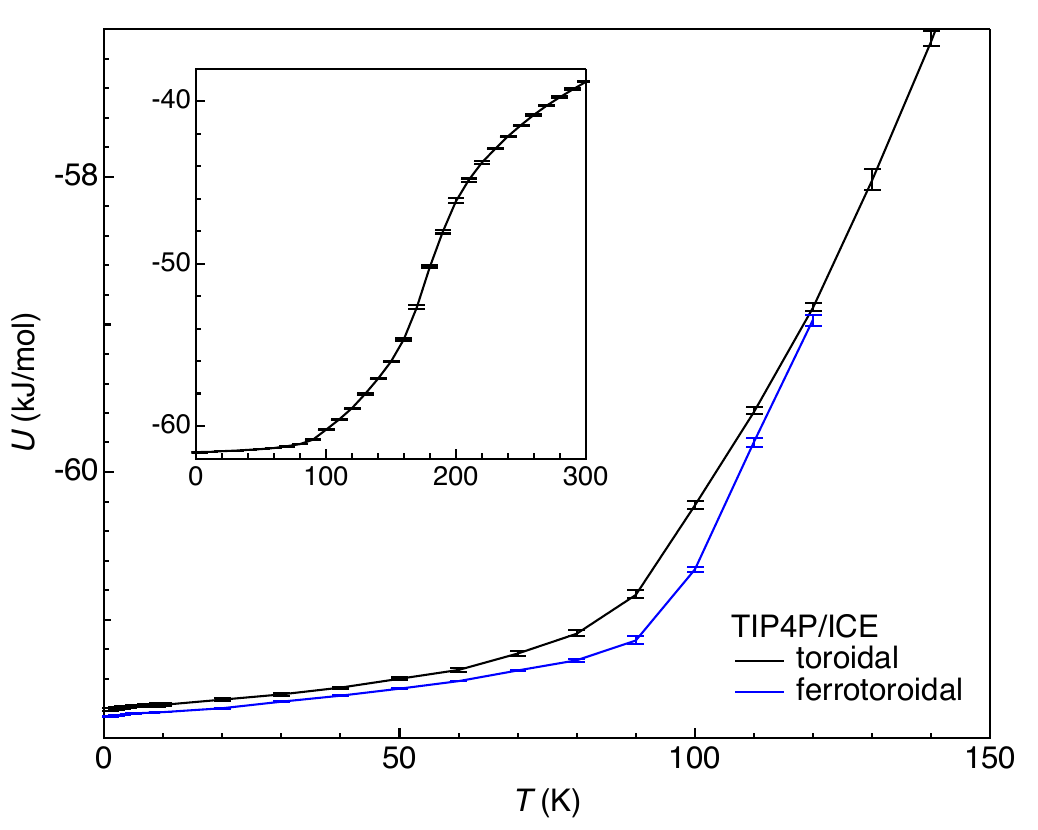}
\caption{\label{fig:potential}
Potential energies of the systems with ferrotoroidal ice and toroidal ice structures.
The initial configuration of the ferrotoroidal ice was artificially constructed, while that of the toroidal ice was that which spontaneously formed during the cooling process of the MD simulation. The error bars are shown at each temperature. The inset shows the results from 300 K to 0 K.} 
\end{figure}


\begin{figure}[tb]
\centering
\includegraphics[width=8.5cm]{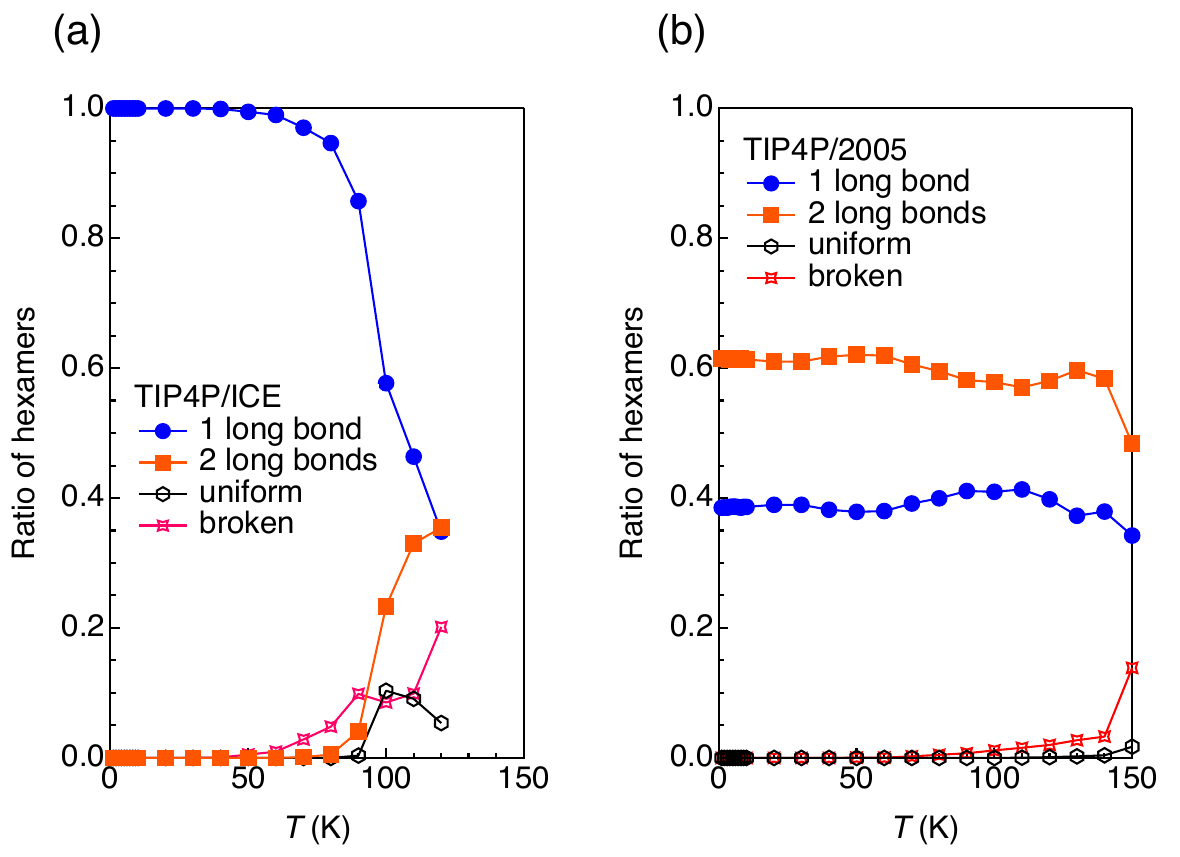}
\caption{\label{fig:longhex}
Ratio of the deformed hexamers as a function of temperature.
The results using (a) TIP4P/ICE and (b) TIP4P/2005 models are shown. 
} 
\end{figure}

Both of the water models give qualitatively the same results for the hexamer formation by hydrogen bonding. 
Figure S\ref{fig:n_hb} shows the number $n_\text{HB}$ of hydrogen bonds per water molecule as a function of temperature. 
In the case of TIP4P/2005, $n_\text{HB}$, which is slightly smaller at 300 K, increases most rapidly in the range of 190 to 200 K, which is about 20 K higher than in the case of TIP4P/ICE.

When the temperature is lowered from 300 K to low temperatures, e.g. 10 K, the ferrotoroidal order of H$_2$O does not spontaneously appear. 
However, the potential energy for the ferrotoroidal structure (prepared by design and then equilibrated at low temperatures) is lower than for the toroidal structure, as shown in Figure S\ref{fig:potential}. 
This is also the case for TIP4P/2005.

Figure S\ref{fig:longhex} shows the ratio of deformed hexamers in the ferrotoroidal structure as a function of temperature. In the case of TIP4P/ICE, all the hexamers have the same shape with one long bond at low temperatures. 
In the case of TIP4P/2005, however, 60 \% of the hexamers have the shape with two long bonds and 40 \% have the same shape with one long bond. 
Because of the coexisting hexamers with one- and two-long bonds, the octadecamer formation is hindered with the TIP4P/2005 model.
Therefore, the shapes of the hexamers and their distributions are sensitive to the choice of the water model.

\subsection{Correspondence between superlattice peak intensity and hexamer number}

\begin{figure}
\centering
\includegraphics[width=0.9\linewidth]{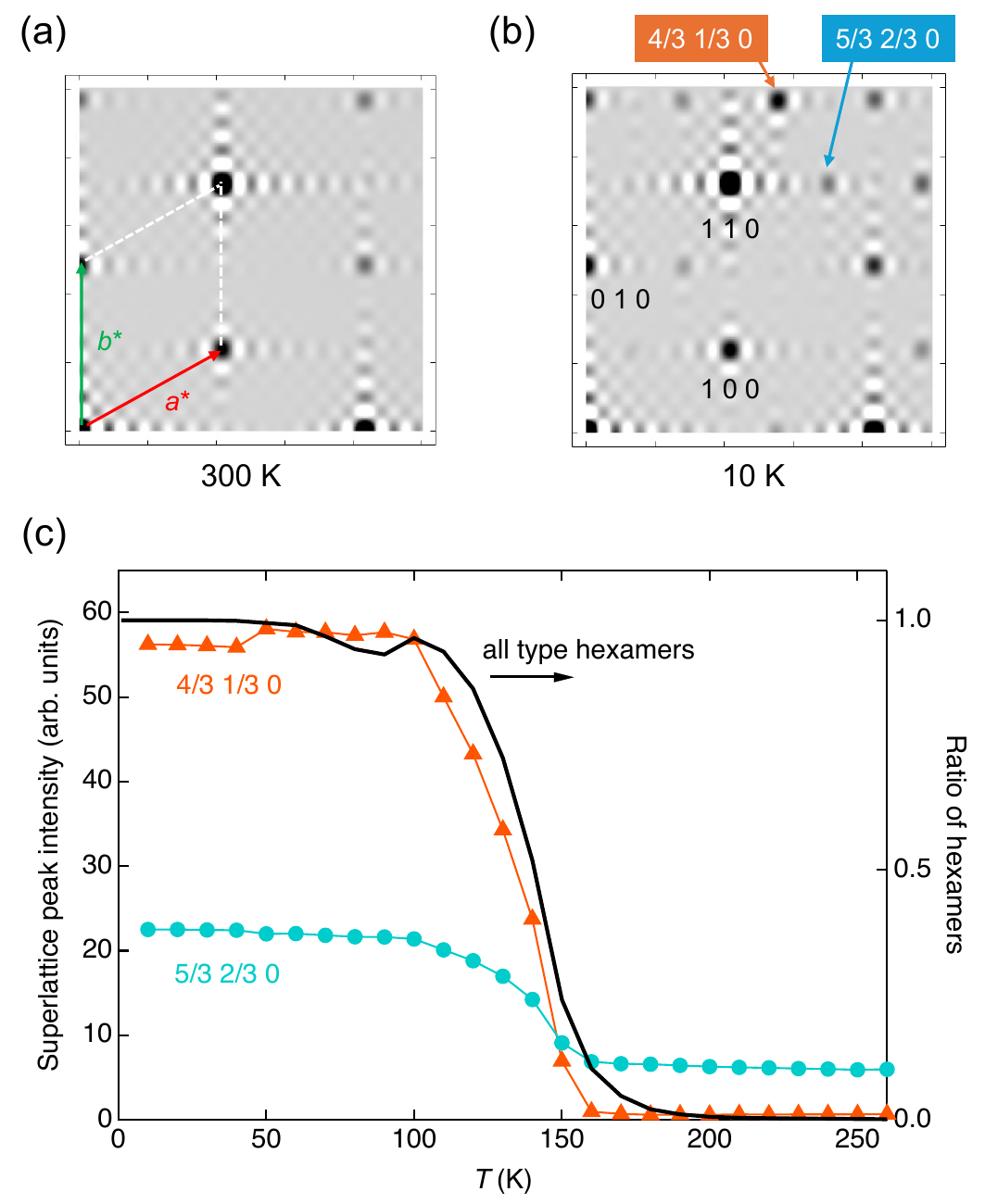}
\caption{\label{fig:super}
(a, b) Two-dimensional structure factor at (a) 300~K and (b) 10~K. 
(c) The peak intensities at the indices of 4/3 1/3 0 (orange) and 5/3 2/3 0 (cyan) are plotted as a function of temperature.
The black curve (right axis) shows the ratio of hexamers, including all types of deformed ones.
} 
\end{figure}

We calculated the two-dimensional structure factor of O atoms of water in martyite by evaluating
$$
S(\mathbf{k}) = 1 + \left<\frac{1}{N}\sum_{i=1}^N \sum_{j\neq i}^N \exp[-i \mathbf{k}\cdot (\mathbf{r}_i-\mathbf{r}_j)]  \right>,
$$
where $\mathbf{r}$ and $\mathbf{k}$ are the position and reciprocal space vectors in two dimensions. 
Figure~\ref{fig:super} shows the (a) high- and (b) low-temperature $S(\mathbf{k})$ in two dimensions. 
At high temperatures, water molecules cannot form stable hydrogen bonds with each other (Fig.~4(d), 300 K), and therefore their average positions constitute the fundamental honeycomb lattice. 
$S(\mathbf{k})$ at 300~K in Fig.~\ref{fig:super}(a) shows only Bragg peaks. 
However, at sufficiently low temperatures, water molecules are so hydrogen-bonded with each other that they form nothing but hexamers (Fig.~\ref{fig:longhex}), and superlattice peaks are observed. 
Figure~\ref{fig:super}(b) shows the superlattice peaks in $S(\mathbf{k})$ at 10~K.
Plotted in Fig.~\ref{fig:super}(c) are the simulated superlattice peak intensities and the ratio of hexamers, both as a function of $T$. 
It is clearly shown that the peak intensities and the hexamer ratio correspond to each other: as $T$ is decreased, both increase rapidly starting at 150~K and plateau at 100~K. 
Therefore, we conclude that the onset of the superlattice peaks results from the formation of hexamers of hydrogen-bonded water molecules.

\section{Critical temperatures}

We comment on the slight discrepancy of $T_\mathrm{c1}$ observed in the dielectric (170--200~K) and XRD (200~K) experiments.
The primary order parameter of the HT-LT phase transition is the electric toroidal dipole moment $\mathbf{A}$, which is not a polar multipole.
Therefore, the dielectric measurement does not necessarily detect this ordering.
The dielectric anomalies observed at $T_\mathrm{c1}$ (Fig.~\ref{dielectric}) probably reflect the ferrotoroidal domain dynamics, since the domain boundaries can locally possess electric dipole moments \cite{Liu2023}.
Similarly, the deformed hexamers below $T_\mathrm{c2}$ possess electric dipole moments, and the alignment of the type-1 hexamers near the domain boundaries can be influenced by external electric fields.
The frequency dependences of these dielectric anomalies ($T_\mathrm{c1}$ and $T_\mathrm{c2}$) indicate the energy barrier to move these domain boundaries.
Therefore, the thermodynamical transition temperatures are expected to be located at the beginning of the dielectric relaxations, namely, $T_\mathrm{c1}\approx 200$~K and $T_\mathrm{c2}\approx 50$~K.

\end{document}